\documentclass[aps,superscriptaddress,nofootinbib,floatfix,epfs,preprintnumbers]{revtex4}
\usepackage{amsfonts,amscd,amsmath,amssymb,graphicx,color,float}
\usepackage[section]{placeins}
\def\ep{\text{e}}
\def\g{\mathsf{g}}
\def\oh{\frac{1}{2}}
\def\s{\mathsf{s}}
\def\m{\mathsf{m}}
\def\k{\mathsf{k}}
\def\r{\mathsf{r}}
\def\rh{r_h}

\def\rq{r_q}
\def\rv{r_v}
\def\rqb{r_{\bar q}}
\def\QQb{\text{\tiny Q}\bar{\text{\tiny Q}}}
\def\Qqb{\text{\tiny Q}\bar{\text{\tiny q}}}
\def\qQb{\text{\tiny q}\bar{\text{\tiny Q}}}
\def\Qqq{\text{\tiny Qqq}}
\def\Qqqb{\bar{\text{\tiny Q}}\bar{\text{\tiny q}}\bar{\text{\tiny q}}}
\def\Q{\text{\tiny Q}}
\def\Qb{\bar{\text{\tiny Q}}}
\def\est{\sigma_\text{\tiny eff}}
\def\rn{r_{\text{\tiny 0}}}
\begin{document}
\preprint{LMU-ASC 09/20}
\title{String Breaking, Baryons, Medium, and Gauge/String Duality}
\author{Oleg Andreev}
 \affiliation{L.D. Landau Institute for Theoretical Physics, Kosygina 2, 119334 Moscow, Russia}
\affiliation{Arnold Sommerfeld Center for Theoretical Physics, LMU-M\"unchen, Theresienstrasse 37, 80333 M\"unchen, Germany}
%\date{}
\begin{abstract} 
The string breaking phenomenon in QCD can be studied using the gauge/string duality. In this approach, one can make estimates of some of the string breaking distances at non-zero temperature and baryon chemical potential. These point towards the enhancement of baryon production in strong decays of heavy mesons in dense baryonic medium. 
\end{abstract}
%\pacs{empty}
\maketitle
%__________________________________________________________ section 2
\section{Introduction}
\renewcommand{\theequation}{1.\arabic{equation}}
\setcounter{equation}{0}

In the string models the strong decay of hadrons is described through light quark-antiquark pair creation \cite{strings}. The most well-known example is that of a heavy meson decay into a pair of heavy-light mesons

\begin{equation}\label{mmode}
	Q\bar Q\rightarrow Q\bar q+\bar Q q
	\,.
\end{equation}
In fact, this is one of the possible decay modes of the heavy meson - the meson mode. From the string theory viewpoint, one can interpret it as a string rearrangement between the heavy and light sea quarks: $Q\bar Q+q\bar q\rightarrow Q\bar q+q\bar Q$. Although it is true that the meson mode is dominant in the vacuum, there are other decay modes which might be of interest for the physics of strong interactions. The next, in number of light quarks, is the baryon mode 

\begin{equation}\label{bmode}
	Q\bar Q\rightarrow Qqq+\bar Q\bar q\bar q
	\,.
\end{equation}
It is natural to expect that this mode is sub-dominant to the meson mode because it is not energetically favorable, and the probability for a string rearrangement between six quarks is lower than between four unless the sea quarks are regarded as a diquark-antidiquark pair $[qq][\bar q\bar q]$ \cite{diquarks}. If so, then the string rearrangement may occur in the same way as it does in the case of the meson mode.

The presence of baryonic medium adds a crucial twist to the story, as light quarks are now not only due to pair creation but also due to the medium. This is the reason for which a decay mode with three light quarks is allowed

\begin{equation}\label{mbmode}
	Q\bar Q\rightarrow Qqq+\bar Qq
	\,.
\end{equation}
We will call it the meson-baryon mode. From the string theory viewpoint, it can be thought of as a string rearrangement between the heavy quarks and light quarks from the medium: $Q\bar Q+q[qq]\rightarrow Q[qq]+\bar Q q$. In that case a baryon is regarded as a two-body system composed of a quark and a diquark \cite{diquarks}. 

Apart from the string models, the string breaking phenomenon has been studied by lattice gauge theory simulations \cite{latrev}. This provides reliable results, but limited to the results for the meson mode at zero temperature and zero chemical potential. For our purposes, what we need to know from this approach can be summarized as follows. One particularly useful model is that of \cite{drum} which includes a mixing analysis based on a correlation matrix whose elements give rise to a model Hamiltonian 

\begin{equation}\label{HD}
{\cal H}(\ell)=
\begin{pmatrix}
E_{\QQb}(\ell) & g \\
g& 2E_{\Qqb} \\
\end{pmatrix}
\,.
\end{equation}
Here $E_{\QQb}(\ell)$ is the energy of two static heavy quark sources separated by distance $\ell$ and connected by a string, $2E_{\Qqb}$ is the energy of a noninteracting pair of heavy-light mesons, and the off-diagonal matrix element $g$ describes the mixing between these two states. The eigenvalues of this model Hamiltonian correspond to the energy levels of a system containing a static quark-antiquark pair. In this way, the lattice data are well-described by a few fit parameters \cite{bali,bulava}. 

For what follows, it is convenient to introduce a characteristic scale of string breaking. Like in \cite{drum,bulava}, we will take 

\begin{equation}\label{deflc}
E_{\QQb}(\ell_c^{(m)})=2E_{\Qqb}
\end{equation}
as a definition and call $\ell_c$ the string breaking distance. Notice that such a definition differs from that of \cite{bali}. However it has a clear meaning in the dual formulation \cite{a-strb}. 

 Certainly, the original model of \cite{drum} can be extended in several ways, for instance, by adding new light flavors, by considering baryon modes, or by extending to finite temperature and non-zero chemical potential. In particular, the free energy matrix containing the decay modes \eqref{mmode}-\eqref{mbmode} is that 

\begin{equation}\label{FD}
{\cal F}(\ell)=
\begin{pmatrix}
\,F_{\QQb}(\ell) & f_1 & f_2 & f_3\\
f_1& F_{\Qqb}+F_{\qQb} &f_{12} & f_{13}\\
f_2 & f_{12} & F_{\Qqq}+F_{\qQb}	 &f_{23} \\
f_3 & f_{13} & f_{23} & F_{\Qqq}+F_{\Qqqb}\,\,
\end{pmatrix}
\,,
\end{equation}
where $F_{\QQb}(\ell)$ is the free energy of two static heavy quark sources separated by distance $\ell$ and connected by a string in the medium. All the remaining diagonal elements are the free energies of noninteracting  heavy-light mesons and baryons. The off-diagonal elements describe the mixing between the states. The lowest eigenvalue of this matrix gives the minimal free energy of a system containing a static quark-antiquark pair in the medium. Of course, the above form of ${\cal F}$ is suggestive enough, without more details on specific situations. If the baryon chemical potential is quite small, the ${\cal F}$ matrix becomes 

\begin{equation}\label{FDT}
{\cal F}(\ell)=
\begin{pmatrix}
F_{\QQb}(\ell) & f_1 & f_3\\
f_1& F_{\Qqb}+F_{\qQb} &f_{13}\\
f_3 & f_{13} & F_{\Qqq}+F_{\Qqqb} \,
\end{pmatrix}
\,,
\end{equation}
because the meson-baryon decay mode has no physical meaning in this case. 

In analogy with what was done above, we introduce several scales by equating $F_{\QQb}$ with the other diagonal elements of the ${\cal F}$ matrix

\begin{equation}\label{lcTm}
F_{\QQb}(\ell_c^{(m)})=F_{\Qqb}+F_{\qQb}
\,,\qquad
F_{\QQb}(\ell_c^{({b'})})=F_{\Qqq}+F_{\qQb}
\,,\qquad
F_{\QQb}(\ell_c^{({b})})=F_{\Qqq}+F_{\Qqqb}
\,.
\end{equation}
The so-defined $\ell_c$'s depend on temperature and chemical potential. 

The purpose of the present paper is to further advance the use of effective string theories in QCD. Here we continue our study of the string breaking phenomenon using the gauge/string duality. This is a detailed and extended version of \cite{a-strb}. The rest of the paper is organized as follows. We begin in Sec.II by setting the framework and recalling some preliminary results. Then, we consider the correlator of two oppositely oriented Polyakov loops and describe static string configurations which make the leading contributions to it in the hadronic phase. In addition, we also comment on some sub-leading configurations and point out those whose relevance increases when approaching the critical line. In Sec.III, a simple but phenomenologically rather successful model is used to illustrate this. We present our estimates of the string breaking distances at non-zero temperature and chemical potential. Finally, we conclude in Sec.IV with a discussion of some open problems. Additional technical details are included in the Appendices.
%__________________________________________________________________
\section{Gauge/string duality and string breaking}
\renewcommand{\theequation}{2.\arabic{equation}}
\setcounter{equation}{0}

Now we will explain how to analyze some aspects of the phenomenon of string breaking in QCD using the gauge/string duality. There have been several papers on this subject in the literature.\footnote{See, for example, the book \cite{book-u} and references therein.} However, here we go another way to describe disconnected string configurations that is a real alternative to probe flavor branes. 

%__________________________
\subsection{Preliminaries}

We start with some preliminary results. We will consider a class of 5-dimensional geometries which is an extension of that of \cite{az1} to finite temperature and baryon chemical potential in the presence of light quarks. All of those represent a charged black hole in an asymptotically $\text{AdS}$ space. The phenomenon of QCD string breaking is modeled by turning on an open string tachyon background which is responsible for light quarks at string endpoints.\footnote{This motivates the usage of the term tachyon. In the present context, its role is however different from that in string theory in ten dimensions, where the open string tachyon is usually associated with instabilities of non-BPS branes rather than fundamental strings.} Thus strings can terminate on them in the interior of five dimensional space. With this, the general form of the background is   

\begin{equation}\label{metric}
ds^2=\ep^{\s r^2}\frac{R^2}{r^2}\Bigl(f(r)dt^2+d\vec x^2+f^{-1}(r)dr^2\Bigr)
\,,
\qquad
{\text A}=\bigl(\text{A}_0(r),0,\dots,0\,\bigr)
\,,
\qquad
{\text T}={\text T}(r)
\,.
\end{equation}
Here $f(r)$ is a blackening factor. It is a decreasing function of $r$ such that $f(0)=1$ and $f(\rh)=0$. The Hawking temperature, which is identified with the temperature of a dual gauge theory, is $T=\tfrac{1}{4\pi}\vert\partial_r f\vert_{r=\rh}$. The $U(1)$ gauge field associated with the baryon charge of quarks obeys the boundary conditions $A_0(0)=\mu$ and $A_0(\rh)=0$, with $\mu$ a baryon chemical potential. $\text{T}$ is a tachyon field.\footnote{We introduce a single scalar field (tachyon), since in what follows we consider only the case of two light quarks of equal mass.} 

The above form has limiting cases which are noteworthy. In the absence of the tachyon field it reduces to a one-parameter deformation of the Reissner-Nordstr\"om solution in Euclidean $\text{AdS}_5$ \cite{chamblin}, with $\s$ a deformation parameter. In the context of AdS/QCD, this type of deformation was first discussed in relation to cold quark matter in \cite{a-cold}.\footnote{For more recent work, see \cite{PC1,PC2,a-screen}.} To further reduce it to a Schwarzschild black hole in a deformed $\text{AdS}$ space, it is enough to turn the gauge field off. The resulting geometry turns out to be of interest, especially for calculating the expectation value of the Polyakov loop \cite{a-pol}. And in the end, letting $T=0$, one arrives at the form which was originally used to successfully model the heavy quark potential \cite{az1}.

To construct string configurations, we need three basic ingredients. The first is a Nambu-Goto string governed by the action 

\begin{equation}\label{NG}
S_{\text{\tiny NG}}=\frac{1}{2\pi\alpha'}\int d^2\xi\,\sqrt{\gamma^{(2)}}
\,,
\end{equation}
where $\gamma$ is an induced metric, $\alpha'$ is a string parameter, and $\xi^i$ are world-sheet coordinates. 

The second is a baryon vertex. In the context of AdS/CFT correspondence it is a five brane \cite{witten}. At leading order in $\alpha'$, the brane dynamics is determined by its world-volume. So the action is  

\begin{equation}\label{vertex-action}
S_{\text{vert}}={\cal T}_5\int d^6\xi\sqrt{\gamma^{(6)}}
\,,
\end{equation} 
where ${\cal T}_5$ is a brane tension and $\xi^i$ are world-volume coordinates. Since the brane is wrapped on an internal space $\mathbf{X}$, from the five-dimensional point of view the vertex looks point-like. We assume that the same holds in respect of AdS/QCD models. If we place all objects at the same fixed point in the internal space, then its detailed structure is not important, except a possible warp factor depending on the radial direction. In \cite{f-bags} it was observed  that an overall warp factor $\ep^{-\s r^2}$ is useful for modeling the equation of state which is a kind of the fuzzy bag model \cite{pis}. Later, it turned out that this warp factor also yields very satisfactory results, when compared to the lattice calculations of the three quark potential \cite{a-3q}. For our purposes, we pick a static gauge $\xi^0=t$ and $\xi^a=\theta^a$, with $\theta^a$ coordinates on $\mathbf{X}$. The action is then 

\begin{equation}\label{baryon-v}
S_{\text{vert}}=\tau_v\int dt \sqrt{f}\,\frac{\ep^{-2\s r^2}}{r}
\,.
\end{equation}
Here $\tau_v$ is a dimensionless parameter such that $\tau_v={\cal T}_5R\,\text{vol}(\mathbf{X})$, where $\text{vol}(\mathbf{X})$ is a volume of $\mathbf{X}$. 

The third ingredient which takes account of light quarks at string endpoints is a tachyon field. So, we add to the world-sheet action 

\begin{equation}\label{Sq0}
S_{\text{q}}=\int d\tau e\,\text{T}
\,,
\end{equation}
which is the usual sigma-model action for strings propagating in a tachyon background. The integral is over a world-sheet boundary parameterized by $\tau$ and $e$ is a boundary metric. In what follows, we consider only the case of a constant tachyon $\text{T}_0$ and world-sheets whose boundaries are lines in the $t$ direction. In this case, the action written in the static gauge is    

\begin{equation}\label{Sq}
S_{\text q}=\m\int dt \frac{\ep^{\frac{\s}{2}r^2}}{r}\sqrt{f}
\,,
\end{equation}
where $\m=R{\text T}_0$. One immediately recognizes it as the action of a point particle of mass ${\text T}_0$ at rest.\footnote{Many aspects of strings with quarks at the ends have been discussed in the literature for years. See, e.g., \cite{bm, bn-book} and references therein.}

Finally, it remains to mention that in the presence of a background gauge field the string endpoints with attached quarks couple to it. So the world-sheet action includes boundary terms which in the case of interest are given by 

\begin{equation}\label{SA}
S_{\text{\tiny A}}=\mp\frac{1}{3}\int dt\, {\text A}_0
\,.
\end{equation} 
The minus and plus signs correspond to a quark and an antiquark. The numerical factor comes from the relation between the chemical potentials of quarks and baryons at $N_c=3$. 

%_______________________________________________________________________________
\subsection{String breaking}

Now consider the correlator of two oppositely oriented Polyakov loops. On the string theory side, its expectation value is given by the world-sheet path integral so that a string world-sheet has the loops for its boundary. In principle, the integral can be evaluated semiclassically with the result 

\begin{equation}\label{LL}
\langle\,L(0)\,L^{\dagger}(\ell)\,\rangle=\sum_n \omega_n\ep^{-S_n}
\,.
\end{equation}
Here $S_n$ is the world-sheet action evaluated on a classical solution (string configuration), $n$ labels the solutions, and $\omega_n$ is a relative weight factor. For static configurations each $S_n$ reduces to $F_n/T$, where $F_n$ is a free energy of the configuration. Importantly for what follows, the $F_n$'s are the diagonal elements of the ${\cal F}$ matrix. Thus, in this formalism, the string breaking distances have a simple meaning, namely that two exponents are equal to each other at $\ell=\ell_c$.

The correlator reduces to an expectation value of a semi-infinite rectangular Wilson loop at $T=\mu=0$. So the above expression becomes

\begin{equation}\label{wilson}
\langle\,W({\cal C})\,\rangle=\sum_n w_n\ep^{-S_n}
\,,
\end{equation}
 where $S_n$ is expressed in terms of an energy of the configuration $E_n$ and a time interval ${\cal T}$ by $S_n=E_n{\cal T}$. The $E_n$'s are the diagonal elements of the model Hamiltonian, in particular $E_{\QQb}$ and $E_{\Qqb}$ are the diagonal elements of \eqref{HD}. 

%____________________________________________ 
\subsubsection{Deeply inside the hadronic phase}

For $(T,\mu)$ far from the critical values, it is natural to assume that the leading contributions to the correlator come from string configurations associated with colorless states and arranged in number of light quarks. In our discussion, we will restrict this number to $4$. For this case the configurations are shown in Figure \ref{lead}, and more details are contained in Appendix B. Here it holds that $r_h\gg r_w$  
%____________________________________________fig-1 
\begin{figure}[ht]
\includegraphics[width=15cm]{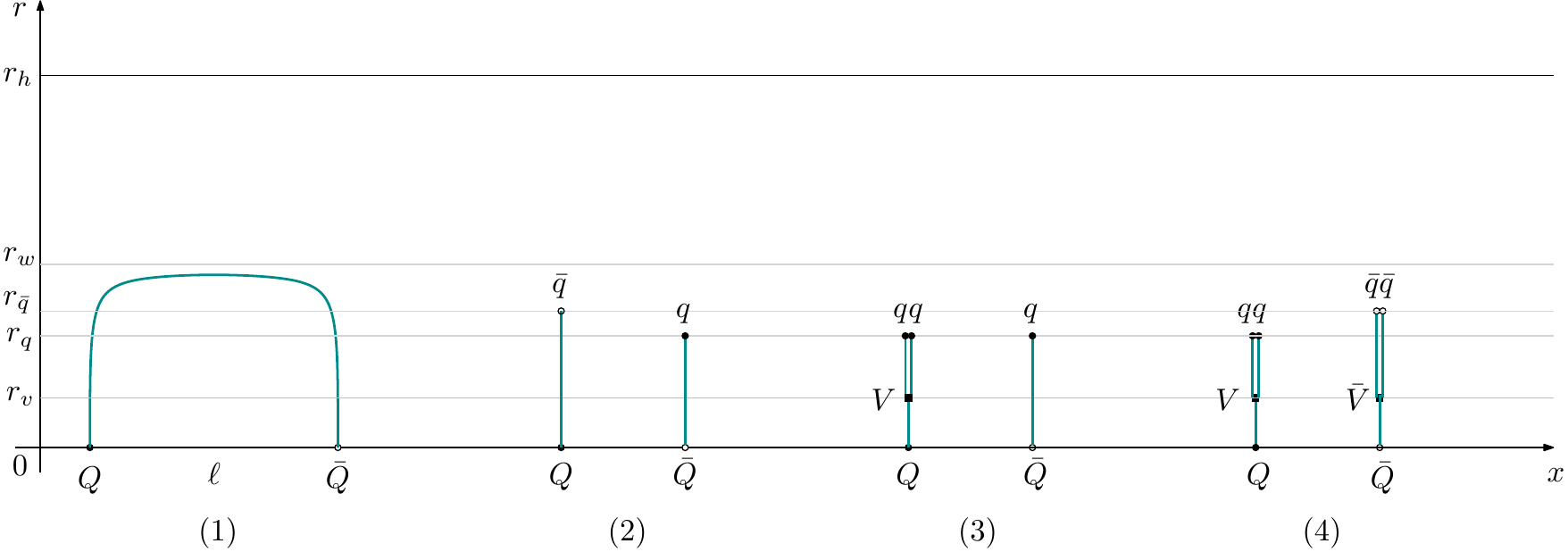}
\caption{{\small The leading string configurations contributing to the Polyakov loop correlator. Heavy (light) quarks are denoted by $Q(q)$, and baryon vertices by $V$. All the strings are in the ground state. The boundary of space is at $r=0$ and the horizon at $r=\rh$. A soft wall responsible for a linear behavior of the free energy of configuration (1) at a large quark separation $\ell$ is located at $r=r_w$. Because the potential $A_0$ gives the electric field in the $r$-direction, the light quarks are shifted towards the boundary.}}
\label{lead} 
\end{figure}
%____________________________________
and $r_{\bar q}>r_q>r_v$. The first is the necessary condition for the model to be in the hadronic phase far from the critical line, where the free energy of configuration (1) behaves linearly as a function of $\ell$ for large $\ell$. The second is the condition of existence of the configurations for the parameter values we use below.  

We begin our discussion with the connected configuration (1) which represents a string stretched between two heavy quark sources. It is clearly that it dominates the Polyakov loop correlator at short distances, where the correlator can be expressed in terms of a single exponential function. However, this comes to naught at larger distances, where the correlator is dominated by disconnected configurations. As a result, the free energy of the pair becomes independent of separation distance (flattened). This is the essence of string breaking. 

 An important fact, explained in Appendix B, is that for large $\ell$ the free energy of this configuration is well approximated by a linear function\footnote{Importantly, for the parameter values we use, this is the case for $\ell\gtrsim 0.5\,\text{fm}$, whereas the string breaking distance is of order $1\,\text{fm}$.}

\begin{equation}\label{FQQb-l}
F_{\QQb}(\ell)=\sigma\ell -2\g\sqrt{\s}\,I+C+o(1)
\,,
\end{equation}
where the string tension $\sigma$ is given by \eqref{sigma} and $I$ by \eqref{c}. $C$ is a normalization constant which is equal to a constant term in the formal expansion of $F_{\QQb}$ near $\ell=0$.

To get further, consider configuration (2). We interpret it as a pair of non-interacting heavy-light mesons. The free energy of the configuration can be directly read off from Eqs.\eqref{FQqb} and \eqref{FQbq}. So, it is 

\begin{equation}\label{Fmeson}
F_{\Qqb}+F_{\qQb}=\sqrt{\s} 
\Bigl(\g{\cal Q}(\bar q)+\m{\cal V}(f;\tfrac{1}{2},\bar q)+(\bar q\rightarrow q)\Bigr)
+\frac{1}{3}
\bigl({\text A}_0(\bar q)-{\text A}_0(q)\bigr)
+C
\,,
\end{equation}
where the functions ${\cal Q}$ and ${\cal V}$ are defined in Appendix A. The positions of the light quarks ($\bar q$ and $q$) are determined from Eqs.\eqref{fb-qb} and \eqref{fb-q}.

Using this expression, we find the string breaking distance 

\begin{equation}\label{lc-mf}
\ell_c^{(m)}=\frac{\sqrt{\s}}{\sigma}
\Bigl(\g{\cal Q}(\bar q)+\m{\cal V}(f;\tfrac{1}{2},\bar q)+(\bar q\rightarrow q)\,
+\frac{1}{3\sqrt{\s}}\bigl({\text A}_0(\bar q)-{\text A}_0(q)\bigr)
+2\g I\,\Bigr)
\,
\end{equation}
that gives a characteristic scale of string breaking when it occurs due to a pair of light quarks. It is noteworthy that the dependence on $C$ cancels out. As a result, $\ell_c^{(m)}$ is a renormalization scheme independent quantity, and therefore is physically meaningful.

Now we consider configuration (3) which represents a non-interacting meson-baryon pair. Since it does not include light antiquarks at the string endpoints, one might think that this configuration is forbidden by baryon number conservation. This is true, except for the baryonic medium, where the light quarks come from the medium and no problem occurs with net baryon number conservation. Thus, this configuration is only meaningful at large enough baryon density (chemical potential).

Using Eqs.\eqref{FQbq} and \eqref{FQqq}, the free energy of the configuration is 
 
\begin{equation}\label{FbaryonI}
F_{\Qqq}+F_{\qQb}=3\sqrt{\s}\Bigl(\g{\cal Q}(q)-\frac{1}{3}\g{\cal Q}(v)
+
\m{\cal V}(f;\tfrac{1}{2},q)
+
\g\k{\cal V}(f;-2,v)
\Bigr)
-{\text A}_0(q)+C
\,,
\end{equation}
where $v$ and $q$ are determined, respectively, from \eqref{fb-v} and \eqref{fb-q}. The corresponding string breaking distance is  

\begin{equation}\label{lc-b1f}
\ell_c^{(b')}=\frac{\sqrt{\s}}{\sigma}
\Bigl(3\g{\cal Q}(q)-\g{\cal Q}(v)+3\m{\cal V}(f;\tfrac{1}{2},q)
+
3\g\k{\cal V}(f;-2,v)
-\frac{1}{\sqrt{\s}}{\text A}_0(q)
+2\g I\,\Bigr)
\,,
\end{equation}
that sets a scale of string breaking for the decay into a meson-baryon pair. 

Unlike configuration (3), another configuration consisting of heavy-light baryons is always meaningful. It is configuration (4). For this configuration, the free energy can be read off from \eqref{FQqq} and \eqref{FQqqb}

\begin{equation}\label{Fbaryon2}
F_{\Qqq}+F_{\Qqqb}=2\sqrt{\s}\Bigl(
\g{\cal Q}(q)+\m{\cal V}(f;\tfrac{1}{2},q)
+(q\rightarrow\bar q)\,
-\g{\cal Q}(v)+3\g\k{\cal V}(f;-2,v)
\Bigr)
+\frac{2}{3}\bigl({\text A}_0(\bar q)-{\text A}_0(q)\bigr)
+C
\,.
\end{equation}
As before, $v$, $\bar q$, and $q$ are the solutions of the equations of Appendix B. The string breaking distance is then 

\begin{equation}\label{lc-b2}
\ell_c^{(b)}=\frac{2\sqrt{\s}}{\sigma}
\Bigl(\g{\cal Q}(q)+\m{\cal V}(f,\tfrac{1}{2},q)
+(q\rightarrow\bar q)\,
-\g{\cal Q}(v)+3\g\k{\cal V}(f;-2,v)
+\frac{1}{3\sqrt{\s}}\bigl({\text A}_0(\bar q)-{\text A}_0(q)\bigr)
+\g I
\Bigr)
\,.
\end{equation}
It gives a characteristic scale when a string is broken by two quark-antiquark pairs which, in the final state, look like a pair of diquarks.\footnote{As seen from the Figure, a diquark is a one-dimensional object extended along the fifth dimension and constructed from two light quarks and a baryon vertex.}

\begin{figure}[ht]
\includegraphics[width=13cm]{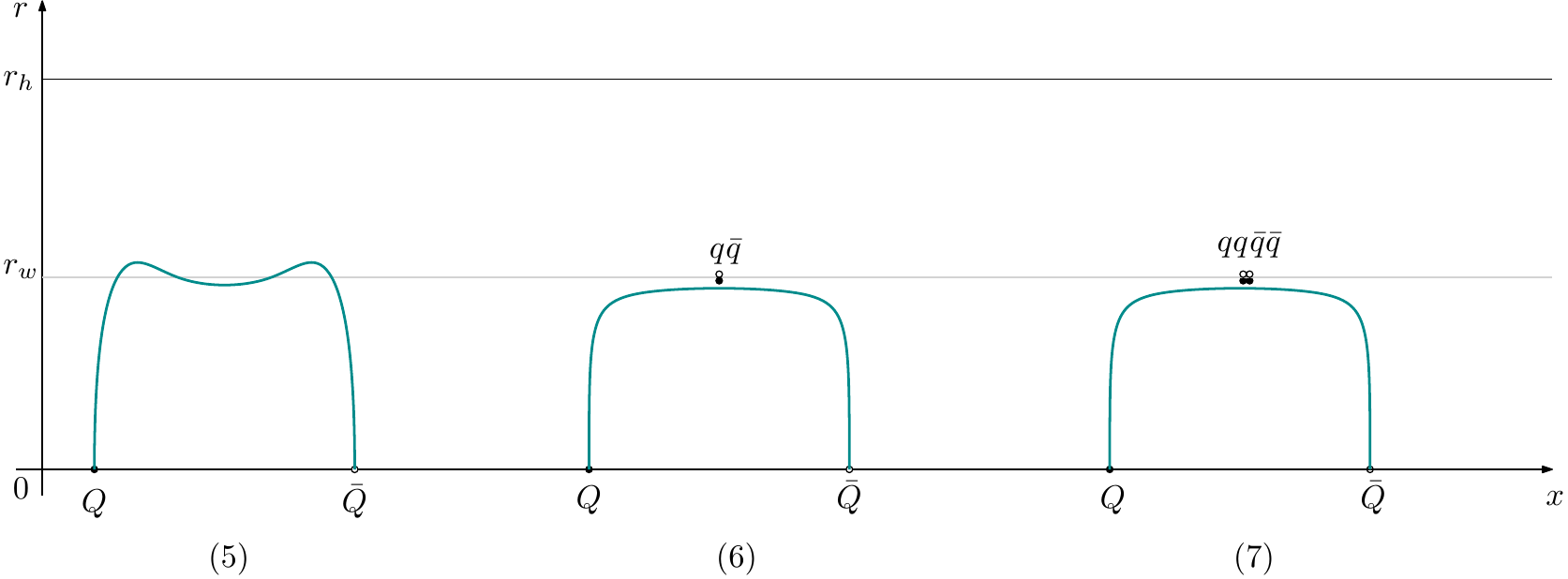}
\caption{{\small Sketched here are the possible subleading configurations. Configuration (5), unlike all the others, includes an excited string. The small circles placed on the soft wall represent bound states of light quarks.}}
\label{sublead} 
\end{figure}

We conclude our discussion with a few remarks on subleading string configurations. Some of those are sketched in Figure \ref{sublead}. The point is that they might  be important in understanding free energies of excited states, but have negligible effect on the ground state free energy.

The connected configuration represents an excited string stretched between the heavy quarks. At zero 
%____________________________________________ fig-2
%____________________________________
temperature and chemical potential such a configuration becomes relevant for describing the hybrid quark potentials \cite{hybrids}. The first disconnected configuration consists of two parts: one is a string stretched between two heavy quarks, and the other is a bound state of two light quarks, say a pion. Thus, its free energy differs from that of configuration (1) by a pion free energy $F_{\pi}$, which reduces to the pion mass at zero temperature and chemical potential. The other disconnected configuration is a generalization of what we have just discussed. Here the pion is replaced by a four quark state, say a sigma meson \footnote{For the present discussion it does not matter how one interprets it either as a tetraquark or as a molecule.}, and therefore $F_\pi$ is replaced by $F_\sigma$ in the expression for the free energy. 

%________________________________________________________________________________
\subsubsection{Near the critical line}
It is easy to admit that near the critical line the hadronic phase contains not only color-singlets but, in addition, some amount of color objects.\footnote{In other words, it becomes a mixed phase.} The simplest of those are quarks that corresponds to a decay 

\begin{equation}\label{Qmode}
	Q\bar Q\rightarrow Q+\bar Q 
	\,.
\end{equation}
It makes sense to explore this in the present context. 

\begin{figure}[ht]
\includegraphics[width=8.25cm]{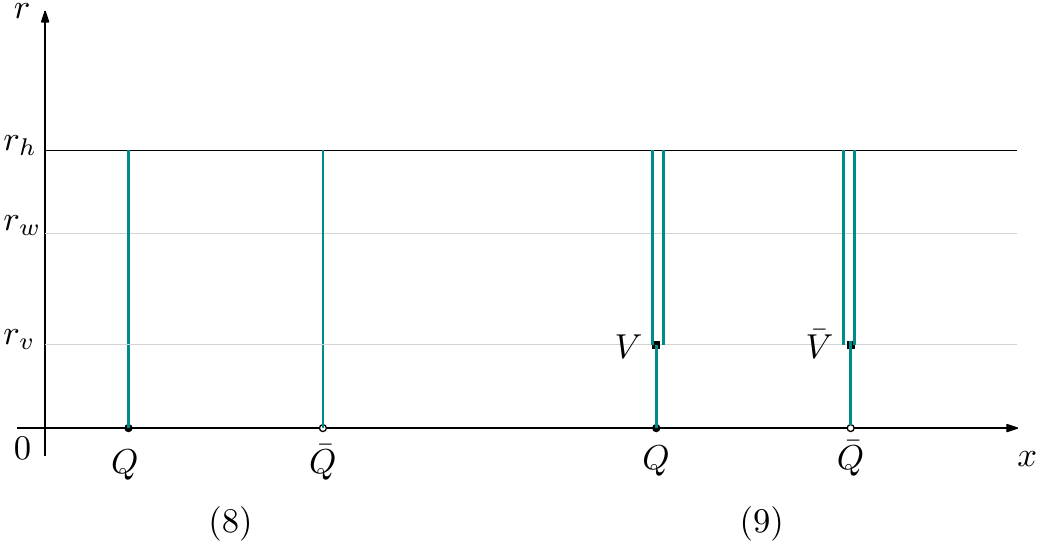}
\caption{{\small The disconnected configurations describing single heavy quarks. They can be interpreted as follows. In the first case the quarks belong to the triplet representation of $SU(3)$, while in the second to the anti-triplet one which is an anti-symmetric two-index representation \cite{a-pol2}.}}
\label{confsQ} 
\end{figure}

As usual in AdS/CFT-like dualities, a static heavy quark is described by a configuration in which a string attached to the quark on the boundary terminates on the horizon (see Appendix B). This means that there are contributions to the Polyakov loop correlator from similar configurations, in particular from those of Figure \ref{confsQ}. In general, it is expected that 
%____________________________________________ fig-3
%____________________________________
a phase transition occurs when the horizon approaches the soft wall \cite{az2} so that temperature and chemical potential reach their critical values. With this in mind, it is intuitively clear that in that case contributions from the configurations of Figure \ref{confsQ} may not be completely negligible with respect to those of Figure \ref{lead}. 

We consider first configuration (8). Combining the expressions \eqref{FQ} and \eqref{FQb}, the free energy of this configuration is simply 

\begin{equation}\label{FQ+FQb}
F_{\Q}+F_{\Qb}
=2\g\sqrt{\s}\,{\cal Q}(h)+C
\,.
\end{equation}
Given the free energy, the corresponding string breaking distance can be defined as before, $F_{\QQb}(\ell_c^{(Q)})=F_{\Q}+F_{\Qb}$. Or more explicitly

\begin{equation}\label{lc-Q}
\ell_c^{(Q)}=\frac{2\g\sqrt{\s}}{\sigma}
\Bigl({\cal Q}(h)+I\,\Bigr)
\,.
\end{equation}

Similarly, for configuration (9), we have 

\begin{equation}\label{FQ+FQb-2}
F_{\Q'}+F_{\Qb'}
=2\g\sqrt{\s}
\Bigl(
2{\cal Q}(h)-{\cal Q}(v)+3\k{\cal V}\bigl(f;-2,v\bigr)
\Bigr)
+C
\,,
\end{equation}
as it follows from the expressions \eqref{FQ-V} and \eqref{FQb-V}. Here $v$ is a solution of  Eq.\eqref{fb-v}. The string breaking distance is then 

\begin{equation}\label{lc-QV}
\ell_c^{(Q')}=\frac{2\g\sqrt{\s}}{\sigma}
\Bigl(
2{\cal Q}(h)-{\cal Q}(v)+3\k{\cal V}\bigl(f;-2,v\bigr)+I\Bigr)
\,.
\end{equation}
The above expressions for the breaking distances give characteristic scales when string breaking results in free quarks. In the case of a pure gauge theory these, however, make no sense because the free energies are infrared divergent in the confined (hadronic) phase.  

%________________________________________________
\section{An Example}
\renewcommand{\theequation}{3.\arabic{equation}}
\setcounter{equation}{0}

To illustrate the above ideas, we consider now a specific model. The good reasons for choosing this model are: (1) Because a string theory dual to QCD is still unknown. It would seem very reasonable to gain experience and intuition by solving problems which can be solved with the effective string model already at our disposal. (2) Because the estimates provided by this model are in agreement with the lattice calculations and QCD phenomenology \cite{az1,a-pol,a-3q,az2}. (3) Because many estimates can be made analytically. (4) Because our goal is to make predictions which may then be tested by means of other non-perturbative methods.

%_________________________________________________
\subsection{The model}

Following \cite{PC1}, we take the blackening factor and gauge field to be of the form

\begin{equation}\label{fnr}
f(r)=1-(1+2p^2)\Bigl(\frac{r}{\rh}\Bigr)^4+2p^2\Bigl(\frac{r}{\rh}\Bigr)^6
\,,\qquad
{\text A}_0(r)=\mu-\r p\frac{r^2}{\rh^3}
\,.
\end{equation}
Here $p$ is a parameter associated with a black hole charge. It takes values on the interval $[0,1]$. $\r$ is a free parameter of the model. Clearly, this background geometry is a simple one-parameter deformation of the Reissner-Nordstr\"om charged black hole in Euclidean $\text{AdS}_5$ \cite{chamblin}, with a deformation parameter $\s$.

Given this, the Hawking temperature and baryon chemical potential, as functions of $p$ and $h$, are  

\begin{equation}\label{Tmu}
T=\frac{1}{\pi}\bigl(1-p^2\bigr)\sqrt{\frac{\s}{h}}
\,,
\qquad
\mu=\r p\sqrt{\frac{\s}{h}}
\,.
\end{equation}
For future convenience we invert these expressions to find 

\begin{equation}\label{hq}
h=\frac{\s\r^2}{\mu^2}\,p^2
\,,
\qquad
p=-\frac{\pi\r}{2}\frac{T}{\mu}+\sqrt{1+\Bigl(\frac{\pi\r}{2}\frac{T}{\mu}\Bigr)^2}
\,.
\end{equation}

%_________________________________________________
\subsection{More details}

In QCD with two light flavors the phase structure is determined by the chiral condensate. In the dual formulation, what is usually regarded as a chiral condensate appears as a coefficient of the subleading term in the expansion of a scalar field near $r=0$ \cite{son}. Setting ${\text T}=const$ is equivalent to a truncation of ${\text T}(r)$ to its leading term. Therefore, we need to assume a working definition which will give us at least a qualitative picture of the phase diagram. Following \cite{az3}, we treat $F_{\QQb}$ as an "order parameter". Such a definition is to some extent model-dependent and motivated by the examples from AdS/QCD \cite{PC1,az3}. The question now is, how accurate is it? As we will see below, the answer to this question is that it is quite good as long as $T$ and $\mu$ are not close to their critical values.\footnote{It is worth noting that a full understanding of the critical behavior of QCD with two light quarks is still not in hand for small enough quark masses (in the chiral limit) \cite{N2}.}

\begin{figure}[ht]
\centering
\includegraphics[width=5cm]{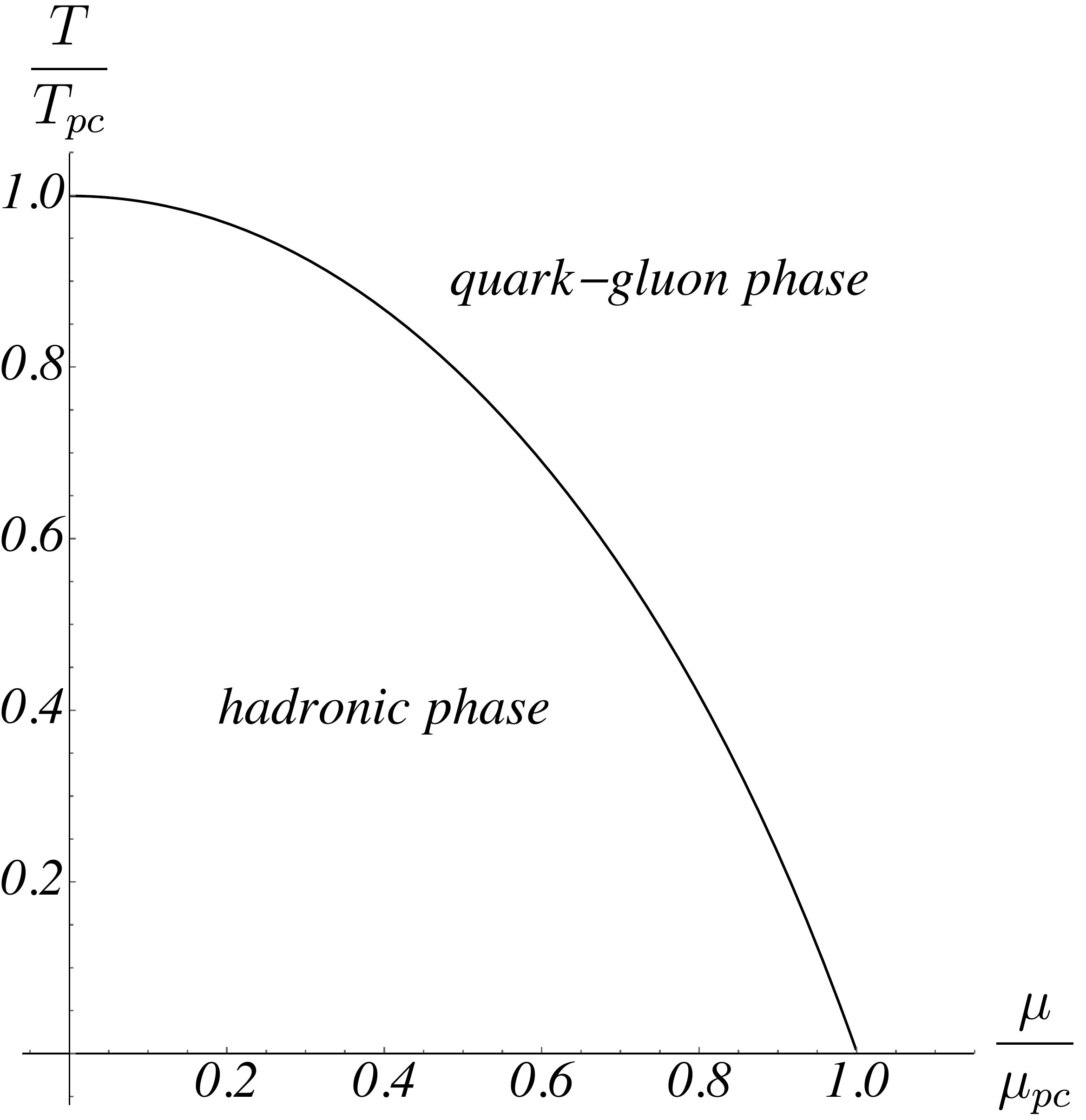}
\caption{{\small The model's phase diagram in the plane of the temperature and baryon chemical potential.}}
\label{phd}
\end{figure}

It follows from the analysis in Appendix C that there are two regimes of behavior for $F_{\QQb}$, one at small values of $T$ and $\mu$ and one at large values. In the former case $F_{\QQb}$ is a linear function of $\ell$ for large $\ell$, while in the latter case not. Therefore, we call them a hadronic phase and a quark-gluon phase, respectively. The phase diagram of the model is shown in Figure \ref{phd}. The phases are separated by the (pseudo) critical line which is determined by Eq.\eqref{critline}. We have 
%________________________  f - 4
%______________________________________________________________  
 \noindent also introduced the naturally-defined pseudocritical temperature and chemical potential

\begin{equation}\label{Tmu-pc}
T_{pc}=\frac{1}{\pi}\sqrt{\frac{\s}{h_0}}
\,,\qquad
\mu_{pc}=\r\sqrt{\frac{\s}{h_1}}
\,,
\end{equation}
where $h_0$ and $h_1$ are given by \eqref{hc0} and \eqref{hc1}. 

For practical use, it is convenient to write the formulas of Sec.II more explicitly by taking into account the expressions \eqref{fnr}. This can be done with the help of a little algebra. So, the force balance equations \eqref{fb-qb} and \eqref{fb-q} take the form 

\begin{equation}\label{fb-Q}
\g\,\ep^{q}+\m\,\ep^{\oh q}
\Bigl(1-(1+2p^2)\bigl(\tfrac{q}{h}\bigr)^2+2p^2\bigl(\tfrac{q}{h}\bigr)^3\Bigr)^{\oh}
\Biggl(q-1+2\bigl(\tfrac{q}{h}\bigr)^2\frac{3p^2\tfrac{q}{h}-1-2p^2}{1-(1+2p^2)\bigl(\tfrac{q}{h}\bigr)^2+2p^2\bigl(\tfrac{q}{h}\bigr)^3}\Biggr)
=\mp\frac{2}{3}\r p \bigl(\tfrac{q}{h}\bigr)^{\frac{3}{2}}
\,.
\end{equation}
Here the minus sign refers to quarks and the plus sigh to antiquarks, where $q$ is replaced by $\bar q$. At the same time the equation for the force balance at the baryon vertex becomes 

\begin{equation}\label{fb-V}
1+3\k\,\ep^{-3v}
\Bigl(1-(1+2p^2)\bigl(\tfrac{v}{h}\bigr)^2+2p^2\bigl(\tfrac{v}{h}\bigr)^3\Bigr)^{\oh}
\Biggl(1+4v+2\bigl(\tfrac{v}{h}\bigr)^2\frac{1+2p^2-3p^2\tfrac{v}{h}}{1-(1+2p^2)\bigl(\tfrac{v}{h}\bigr)^2+2p^2\bigl(\tfrac{v}{h}\bigr)^3}\Biggr)
=0
\,,
\end{equation}
as follows from \eqref{fb-v}. 

One can do a similar calculation for the string breaking distances, with the result\footnote{The expression for $\ell_c^{(Q)}$ remains unchanged.}

\begin{gather}
\ell_c^{(m)}=\frac{\sqrt{\s}}{\sigma}
\Bigl(\g{\cal Q}(\bar q)+\m{\cal V}\bigl(p,\tfrac{\bar q}{h};\tfrac{1}{2},\bar q\bigr)
+(\bar q\rightarrow q)
+
\frac{1}{3}\r\, p h^{-\frac{3}{2}}\bigl(q-\bar q\bigr)
+2\g I\,\Bigr)
\,, \label{lc-m}\\
\ell_c^{(b')}=\frac{\sqrt{\s}}{\sigma}\Bigl(3\g{\cal Q}(q)-\g{\cal Q}(v)
+
3\m{\cal V}\bigl(p,\tfrac{q}{h};\tfrac{1}{2},q\bigr)
+
3\g\k{\cal V}\bigl(p,\tfrac{v}{h};-2,v\bigr)+\r ph^{-\frac{3}{2}}q
-\tfrac{\mu}{\sqrt{\s}}+2\g I\,\Bigr)
\,,\label{lc-b1}\\
\ell_c^{(b)}=\frac{2\sqrt{\s}}{\sigma}
\Bigl(
\g{\cal Q}(q)+\m{\cal V}\bigl(p,\tfrac{q}{h};\tfrac{1}{2},q\bigr)
+(q\rightarrow\bar q)
-\g{\cal Q}(v)+3\g\k{\cal V}\bigl(p,\tfrac{v}{h};-2,v\bigr)
+\frac{1}{3}\r\, p h^{-\frac{3}{2}}\bigl(q-\bar q\bigr)
+\g I\,\Bigr)
\,,\label{lc-Q'}\\
\ell_c^{(Q')}=\frac{2\g\sqrt{\s}}{\sigma}
\Bigl(
2{\cal Q}(h)-{\cal Q}(v)+3\k{\cal V}\bigl(p,\tfrac{v}{h};-2,v\bigr)
+I\Bigr)
\,.\label{lc-QV2}
\end{gather}
The light quark and vertex positions are determined from \eqref{fb-Q}-\eqref{fb-V} and therefore these expressions depend only on the parameters $p$ and $h$. The last, in turn, are expressed in terms of $T$ and $\mu$ by means of \eqref{hq}. 
%_________________________________________________________________________
\subsection{Numerics}

It is of great interest to see how the string breaking distances behave as temperature and chemical potential are varied. To this end, we first need to fix the free parameters of the model.\footnote{In fact, we need to fix only one parameter which is $\m$. All the others have already been fixed in previous studies conducted in the context of this model.} This can be done in two different ways \cite{a-strb}. The first way is to mainly use the results of lattice QCD available at zero temperature and baryon chemical potential. In doing so, the value of $\s$ is fixed from the slope of the Regge trajectory of $\rho(n)$ mesons in the soft wall model with the geometry \eqref{metric}. This gives $\s=0.450\,\text{GeV}^2$ \cite{a-q2}. Then, using \eqref{I0}, we obtain $\g=0.176$ by fitting the value of the string tension $\sigma_0$ to its value in \cite{bulava}. The parameter $\m$ is adjusted to reproduce the lattice result for the string breaking distance $\ell_c^{(m)}$. With $\ell_c^{(m)}=1.22\,\text{fm}$ \cite{bulava}, this gives $\m=0.538$. In \cite{a-3q}, the value of $\k$ is adjusted to fit the three-quark potential to the lattice data for pure $SU(3)$ gauge theory. So far there is no such data available for QCD with two dynamical quarks. We take $\k=-\tfrac{1}{4}\ep^{\frac{1}{4}}$ simply because it yields an exact solution to Eq.\eqref{fb-V}, namely $v=\tfrac{1}{12}$.\footnote{If $T=\mu=0$, then a simple analysis shows that on the interval $[0,0.566]$, with the upper limit being a solution of \eqref{fb-Q} at $\m=0.538$, equation \eqref{fb-V} has solutions if $-0.558\lesssim \k\leq -\tfrac{1}{4}\ep^{\frac{1}{4}}$.} For a shorthand, we denote this set of parameter values by $L$. 

We are now in a position to present some simple estimates obtained for the parameter set $L$. At zero temperature and chemical potential, we get \cite{a-strb} 

\begin{equation}\label{lcb-lattice}
\ell^{(b)}_c=2.35\,\text{fm}
\,.
\end{equation}
Thus the baryon mode is sub-dominant, as expected. Note that for the allowed values of $\k$, $\ell_c^{(b)}$ is a slowly varying function of $\k$ which can take values from $2.23\,\text{fm}$ to $2.35\,\text{fm}$. So the error associated with our choice of $\k$ is, in fact, less than $5\%$. As for the pseudocritical temperature and chemical potential \eqref{Tmu-pc}, their values are given by $T_{pc}=132\,\text{MeV}$ and $\mu_{pc}=538\,\text{MeV}$. 

In \cite{bulava} the numerical calculations were done at unphysical pion mass $m_{\pi}=280\,\text{MeV}$. In this light and in view of possible applications to phenomenology, we now consider the second way of fixing the parameters. In this way, the values of $\s$ and $\g$ are extracted from the quarkonium spectrum obtained by using the heavy quark potential derived from the model \cite{az1}. This is self-consistent, and gives $\s=0.15\,\text{GeV}^2$ and $\g=0.44$ at $T=\mu=0$ \cite{giannuzzi}. The value of $\k$ is set to $-\tfrac{1}{4}\ep^{\frac{1}{4}}$, as before. We determine $\m$ from the condition $E_{\Qqq}-E_{\Qqb}=M_{\Lambda_c^+}-M_{D^0}\approx 420\,\text{MeV}$ \cite{pdg}. This results in $\m=0.699$. We denote this set by $P$. Using it, we find that 

\begin{equation}\label{lc=pheno}
\ell_c^{(m)}=1.07\,\text{fm}\,,
\qquad
\ell_c^{(b)}=1.99\,\text{fm}
\,
\end{equation}
at $T=\mu=0$. These values are smaller than those above, that could correspond to a more physical situation, with a lighter pion. In this case, the pseudocritical temperature and chemical potential turn out to be about $76\,\text{MeV}$ and $621\,\text{MeV}$, respectively. 

%________________________  f - 5 
\begin{figure}[bh]
\centering
\includegraphics[width=6.35cm]{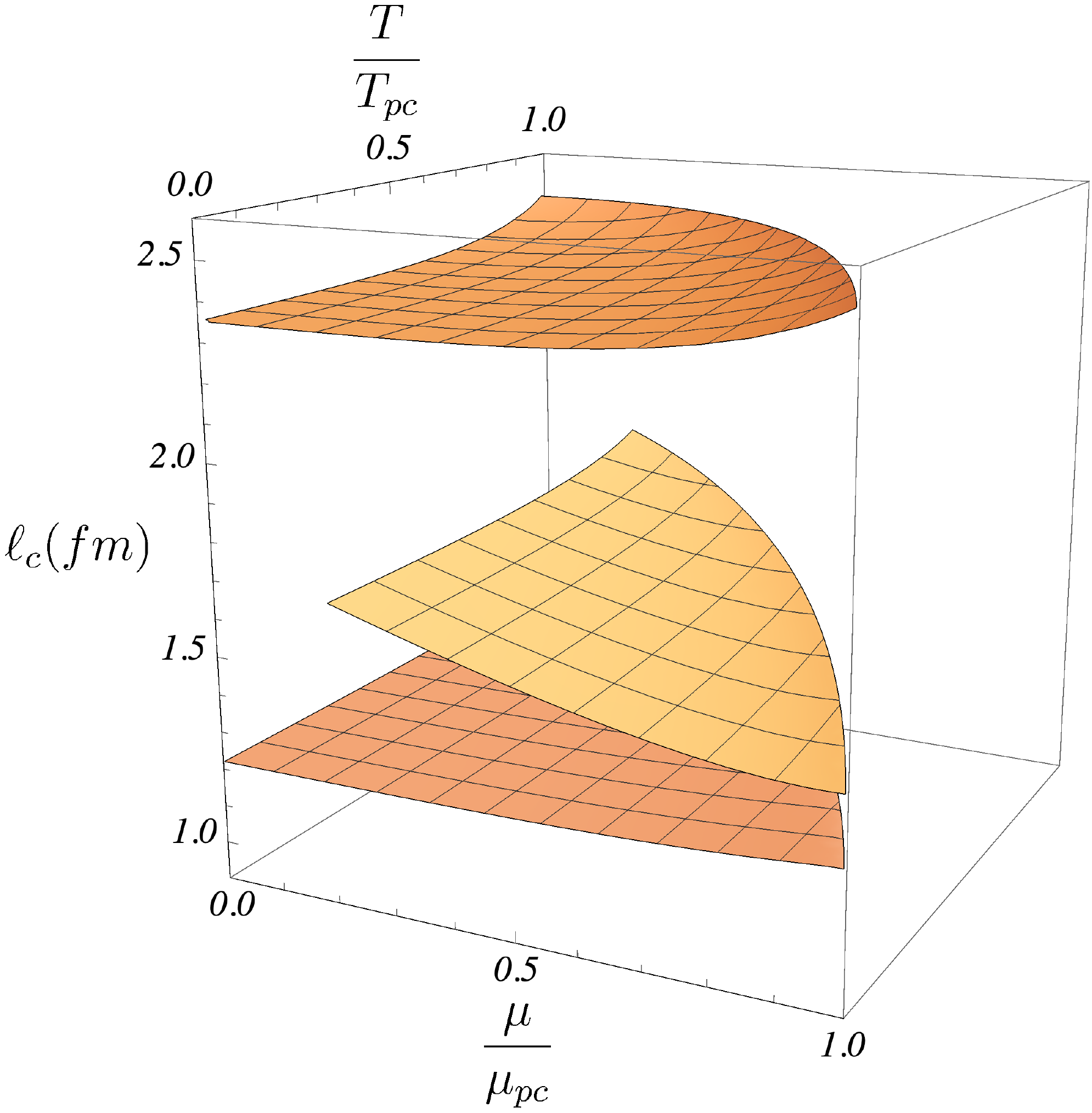}
\hspace{2.25cm}
\includegraphics[width=6.25cm]{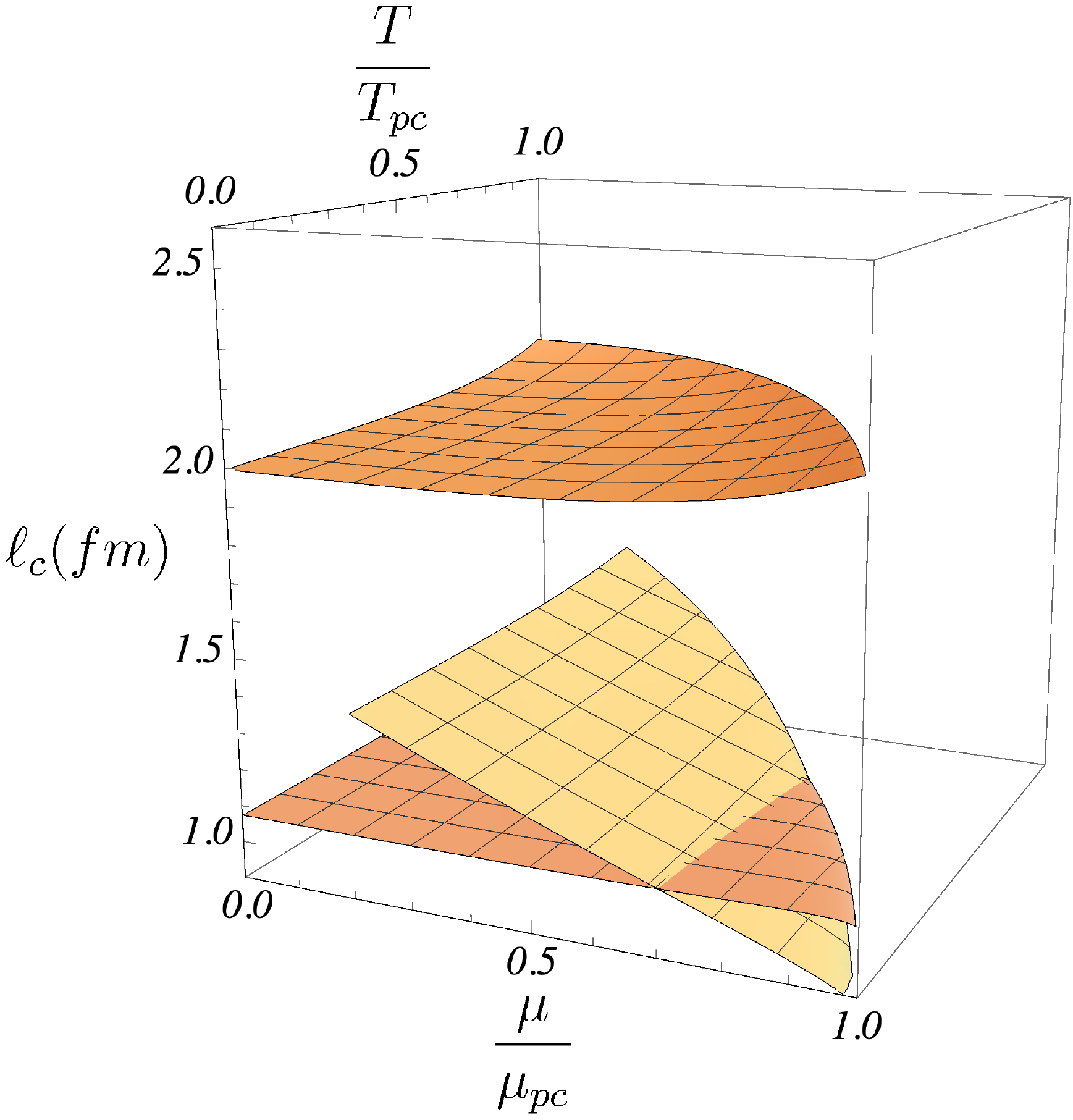}
\caption{{\small The string breaking distances $\ell_c^{(m)}$, $\ell_c^{(b')}$, and $\ell_c^{(b)}$ (shown from bottom to top). For $\ell_c^{(b')}$ the chemical potential runs from $0.2\mu_{pc}$ to $\mu_{pc}$ because this distance is meaningful only if the chemical potential is different enough from zero. The left panel corresponds to the set $L$ with $\r=1.5$, and the right panel to the set $P$ with $\r=3$.}}
\label{lc}
\end{figure}
%______________________________________________________________

In each of these cases, we still need a value of $\r$ to actually make the estimates of the string breaking distances at non-zero chemical potential. As before, one way would be to use lattice results, but now with the caveat that those are available only at small baryon chemical potential. On the other hand, the estimates of the Debye screening mass, which are qualitatively consistent with the lattice, point out that $\r$ is somewhere between $2$ and $6$ \cite{a-screen}. In Figure \ref{lc}, we present our results. We see that the baryon mode is always sub-dominant, as one could expect. Its string breaking distance is roughly twice the distance for the meson mode. Both $\ell_c^{(m)}$ and $\ell_c^{(b)}$ are slowly increasing functions of temperature and chemical potential. By contrast, the string breaking distance $\ell_c^{(b')}$ is slowly increasing with temperature but noticeably decreasing with chemical potential. Thus the energetic preference of the meson decay mode over the meson-baryon one decreases with the increase of baryon chemical potential (baryonic density). Moreover, the meson-baryon mode might even become dominant for some values of the parameters, as seen from the right panel in the Figure above. 

\begin{figure}[ht]
\centering
\includegraphics[width=6.3cm]{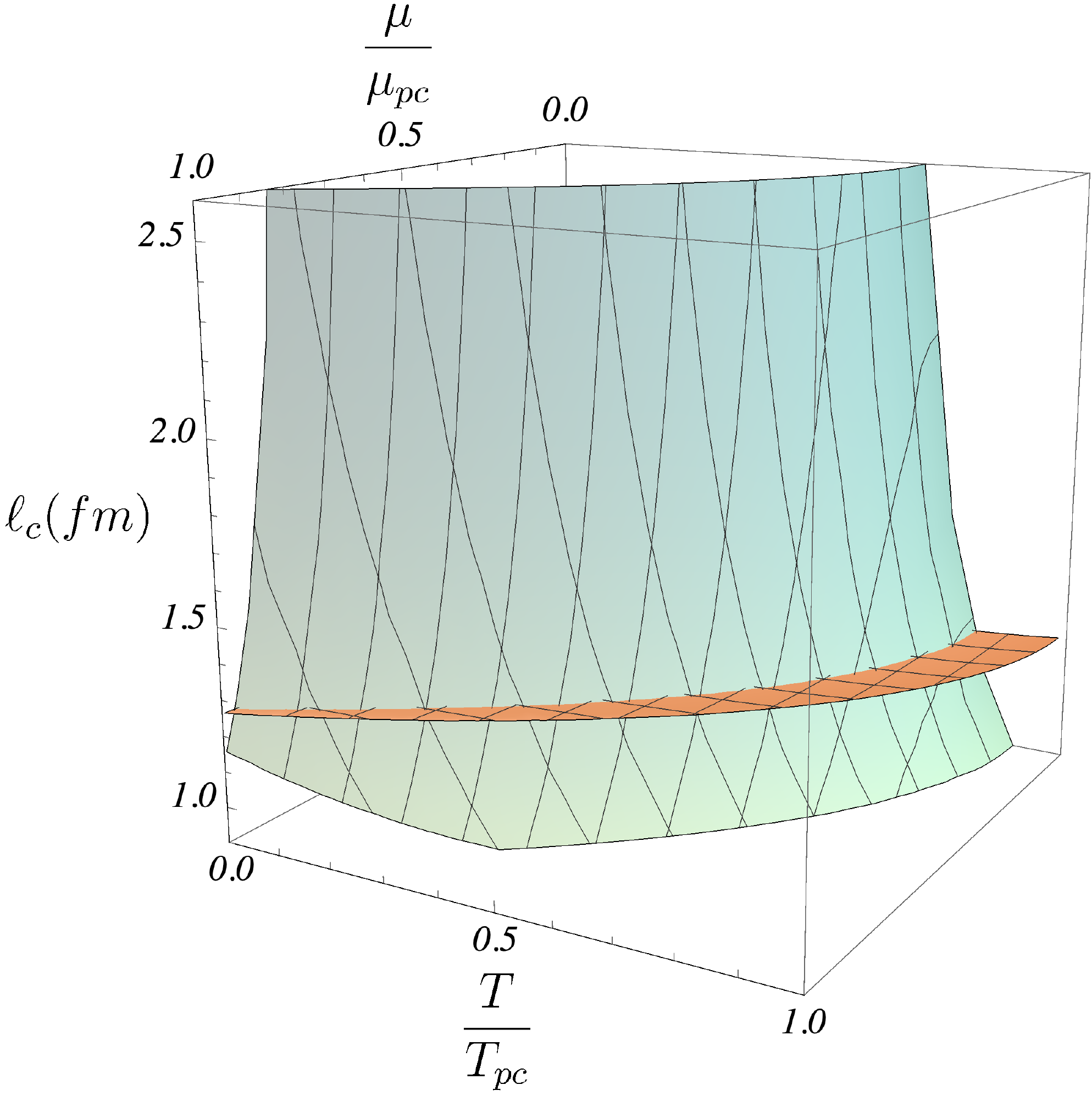}
\hspace{2.25cm}
\includegraphics[width=6.25cm]{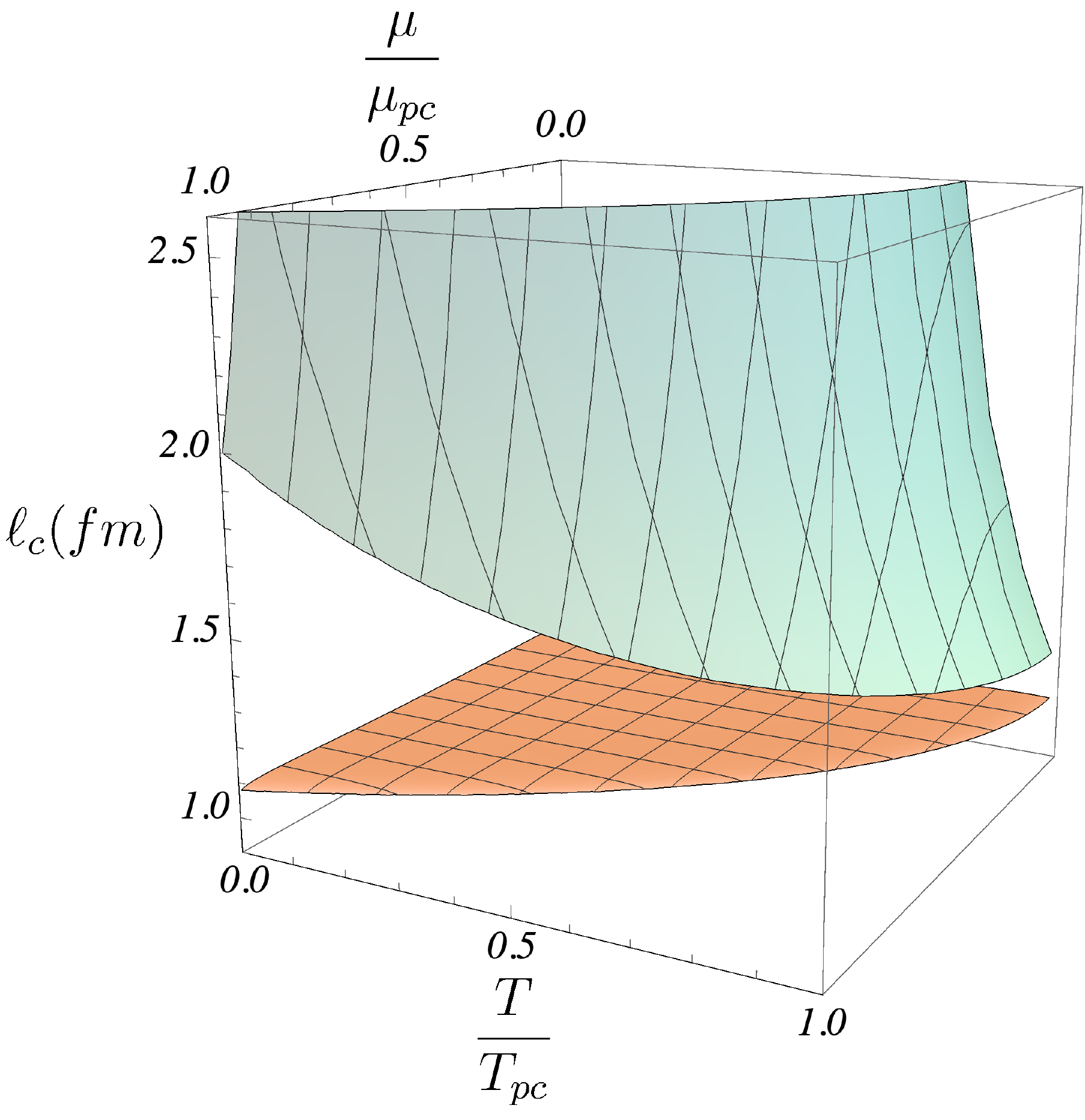}
\caption{{\small The string breaking distances $\ell_c^{(m)}$ and $\ell_c^{(Q)}$ (shown in light cyan). As above, the left panel refers to the parameter set $L$  with $\r=1.5$, and the right panel to the set $P$ with $\r=3$.}}
\label{lcQ}
\end{figure}

There is certainly something to be said about the subleading string configurations of Figure \ref{sublead}, but we will leave this for future work. We just conclude with a few comments on the disconnected configurations shown in Figure \ref{confsQ}. In the context of AdS/QCD these describe single quarks. In Figure \ref{lcQ}, we present the results for the string breaking %________________________  f - 6 __________________________________
%______________________________________________________________
\noindent distance $\ell_c^{(Q)}$ and their comparison with those for $\ell_c^{(m)}$. Obviously, $\ell_c^{(Q)}$ shows a steep fall between the origin, where it is singular, and the critical line, where it becomes comparable with the distance $\ell_c^{(m)}$. The physical meaning of such behavior is easily understood by recalling that the free energy of a single quark is IR divergent at $T=\mu=0$ and a number of free quarks vastly grows as $T$ and $\mu$ tend to the critical values.  To simplify the Figure, we do not explicitly show the results for the string breaking distance $\ell_c^{(Q')}$. The reason is that although the function $\ell_c^{(Q')}$ behaves similarly to $\ell_c^{(Q)}$, it takes larger values. Thus, it turns out that the string configuration (9) is subleading to (8).\footnote{A caveat here is that this conclusion is valid at least for the parameter sets we use. The difference between the free energies $F_{Q'}$ and $F_{Q}$ might be small, as it was observed in \cite{a-pol2}.}
%________________________________________________________________________
\section{Concluding Comments}
\renewcommand{\theequation}{4.\arabic{equation}}
\setcounter{equation}{0}

(i) One of the main purposes of the AdS/QCD approach is to make predictions on the strongly coupled regimes of QCD that would be otherwise impossible or difficult to explore using the standard tools. In this regards, the two main conclusions we have drawn in \cite{a-strb} can be summarized as follows. First, the energetic preference of the meson decay mode over the meson-baryon one decreases with chemical potential (baryon density).\footnote{The latter might even become energetically favorable, as seen from the right panel in Figure \ref{lc}. In this case a simple estimate gives $435\,\text{MeV}$ for the transition value of $\mu$.} Combining this with the natural assumption that the probability for a string rearrangement between the heavy quarks and light quarks coming from the medium also increases with an increase of baryon density, it is natural to expect the enhancement of heavy-light baryon production in strong decays of heavy mesons in the dense baryon medium. Second, in the process of hadronization the $\Lambda$-baryons and $D$-mesons are formed from $c$-quark coalescence with light quarks and antiquarks. Assuming that the hadrons are at rest in the plasma frame, we can estimate the difference between the free energies of heavy-light mesons and baryons. Since it decreases with chemical potential, meson formation becomes less and less favorable. This could be one of the reasons for the enhanced $\Lambda_c^+$ production in PbPb collisions (with respect to pp collisions) measured with the ALICE detector at CERN \cite{alice}. Clearly, the same argument that we gave applies to the $b$-quark too. The results of this paper show that in the hadronic phase temperature effects do not alter these conclusions.\footnote{The point is that the shorter the breaking distance, the smaller the free energy.}

(ii) We have discussed the phenomenon of string breaking within the effective string model in the background fields defined by Eq.\eqref{metric}. As an important illustration of these ideas, we gave the example for the specific background. The model we are pursuing has its own limitations and shortcomings, as any model. Apparently, there are many things which deserve to be further clarified and improved. For instance, one of those is a role playing by the open tachyon field and its possible implications on the structure of phase diagram. In another direction, there is a circle of questions related to the string breaking phenomenon which can be addressed using the model already at our disposal. In this regards, it would be particularly interesting to see what happens in the case of triply heavy baryons.

%__________________________________________________________________
\begin{acknowledgments}
This work was supported in part by RFBR Grant 18-02-40069. We are grateful to I.Ya. Aref'eva, P. de Forcrand, A. Francis, A. Vairo, P. Weisz, and U.A. Wiedemann for discussions concerning this topic. We also thank the Arnold Sommerfeld Center for Theoretical Physics and CERN Theory Division for their hospitality. 
\end{acknowledgments}

\appendix
\section{Some useful formulas}
\renewcommand{\theequation}{A.\arabic{equation}}
\setcounter{equation}{0}
%____________________________________________________________________________________

For convenience, we present here a couple of integrals appearing in the calculations. Using integration by parts, the integrals can be simplified and expressed in terms of the imaginary error function, $\text{erfi}(z)=\frac{2}{\sqrt{\pi}}\int_0^z dx\,\ep^{x^2}$. So, 

\begin{equation}\label{Ioa}
\int^a_0\frac{dx}{x^2}\bigl(\ep^{cx^2}-1\bigr)=
\sqrt{\pi c}\,\text{erfi}(a\sqrt{c})	+
\frac{1}{a}\Bigl(1-\ep^{ca^2}\Bigr)
\,
\end{equation}
and
\begin{equation}\label{Iab}
\int_a^b \frac{dx}{x^2}\,\ep^{cx^2}=	\sqrt{\pi c}\Bigl(\text{erfi}(b\sqrt{c})-\text{erfi}(a\sqrt{c})\Bigr)-\frac{\ep^{cb^2}}{b}+\frac{\ep^{ca^2}}{a}
\,,
\end{equation}
where $a,b,$ and $c$ are positive numbers. 

For what follows, it is also convenient to define a function

\begin{equation}\label{Q}
{\cal Q}(x)=\sqrt{\pi}\,\text{erfi}(\sqrt{x})-\frac{\ep^{x}}{\sqrt{x}}
\,
\end{equation}
which allows one to write the results in a simpler form. 

The peculiar forms of the actions \eqref{baryon-v} and \eqref{Sq} motivate the definition 

\begin{equation}\label{Vf}
{\cal V}(f;b,y)=\sqrt{\frac{f}{y}}\,\ep^{by}
\,.
\end{equation}
In particular, it takes the form 

\begin{equation}\label{V}
{\cal V}(a,x;b,y)=\sqrt{1-(1+2a^2)x^2+2a^2x^3}\,\frac{\ep^{by}}{\sqrt{y}}
\,
\end{equation}
for the background geometry \eqref{fnr}. Here $(a,x)$ belong to the interval $[0,1]$.

%_____________________________________________
\section{Basic string configurations}
\begin{figure}[ht]
\includegraphics[width=16cm]{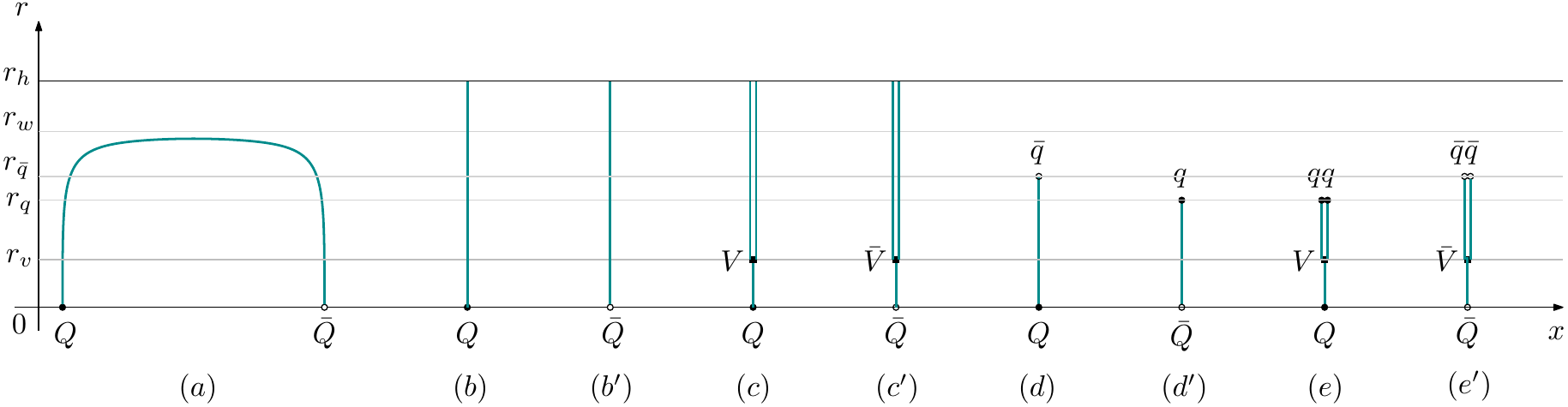}
\caption{{\small Basic string configurations. Here notation is the same as in Figure \ref{lead}.}}
\label{bconfs} 
\end{figure}

%\label{notation}
\renewcommand{\theequation}{B.\arabic{equation}}
\setcounter{equation}{0}
In this Appendix we consider the static string configurations sketched in Figure \ref{bconfs}. This is a basic set which provides a 
%____________________________________________fig - 7
%____________________________________________
proper framework for evaluating the configurations of Sec.II. We arrange the configurations according to increasing number of light quarks at string endpoints. 
%____________________________________________________________________________________
\subsection{Configurations without light quarks}
We begin by briefly summarizing the results on configurations $(a)\text{-}(c')$. In the case of zero chemical potential or vanishing baryon density, these configurations were widely discussed in the literature.\footnote{So, the only new result presented here is that constant terms in the expansions of a free energy of configuration $(a)$ are different for small and large quark separations.}  For standard explanations, see \cite{a-screen,a-pol,a-pol2} whose conventions we generally follow. In the case of finite chemical potential, just a few minor modifications are needed. 

\subsubsection{A connected configuration}
In the gauge/string duality, configuration $(a)$ describes a kind of bound state. It includes two quark sources placed on the boundary of space and a string stretching between them. A gravitational force bends the string by pulling it towards the bulk. This is a crucial difference with a straight string in 4-dimensional models in flat space. Usually one uses this configuration to compute a contribution to the singlet free energy of the quark-antiquark pair. For the background geometry \eqref{metric}, it can be  written in parametric form as \cite{a-screen}

\begin{equation}\label{FQQb}
\ell(\rn)=2\int_0^{\rn} \frac{dr}{\sqrt{f}}\, 
\biggl[\,
\frac{\est^2}{{\mathcal I}^2}-1\,
\biggr]^{-\oh}
\,,
\qquad
F_{\QQb}(\rn)=2\g\int_{0}^{\rn} \frac{dr}{r^2}\biggl(\ep^{\s r^2}
\biggl[\,1-\frac{{\mathcal I}^2}{\est^2}\,\biggr]^{-\oh}-1\biggr)\,\,-\frac{2\g}{\rn}+C
\,,
\end{equation}
Here $\ell$ is a separation distance and $\rn$ is a parameter defined by $\rn=\max r$. It runs from $0$ to $r_w$. The multiplicative factor $\g$ is given by $\g=\frac{R^2}{2\pi\alpha'}$ and $C$ is a normalization constant resulting from the subtraction of a linear divergence (infinite quark mass). The effective string tension $\est$ is defined by 

\begin{equation}\label{est}
	\est(r)=\sqrt{f}\,\frac{\ep^{\s r^2}}{r^2}
	\,.
	\end{equation}
${\mathcal I}$ is a first integral which is set to $\est (\rn)$. 

A simple analysis shows that $\ell(\rn)$ is a monotonically increasing function on the interval $[0, r_w]$, and that it tends to zero as $\rn\rightarrow 0$ and to infinity as $\rn\rightarrow r_w$. In the first limiting case, the asymptotic behavior of the free energy is that 

\begin{equation}\label{FQQb-small}
F_{\QQb}(l)=-\frac{\alpha}{\ell}+C+o(1)
\,,
\end{equation}
with $\alpha=\g (2\pi)^3 \Gamma^{-4}\bigl(\tfrac{1}{4}\bigr)$. In the second case, the behavior of $\ell$ and $F_{\QQb}$ near $\rn=r_w$ is 

\begin{equation}\label{Fl}
\ell(\rn)=-k\ln(r_w -\rn)\,+O(1)
\,,\qquad
F_{\QQb}(\rn)=-k\sigma\ln(r_w -\rn)+O(1)
\,,
\end{equation}
where 
\begin{equation}\label{sigma}
k=2\sqrt{\est/f\est''}(r_w)\,,
\qquad
\sigma=\g\est(r_w)
\,.
\end{equation}
From this, it follows that $F_{\QQb}(\ell)=\sigma\ell+O(1)$, with the string tension $\sigma$. 

To find the constant term in the asymptotic expansion for large $\ell$, consider 

\begin{equation}\label{difference}
	F_{\QQb}-\sigma\ell=2\g\int_{0}^{\rn} \frac{dr}{r^2}\biggl(\ep^{\s r^2}
\biggl[1-\frac{{\mathcal I}^2}{\est^2}\,\biggr]^{-\oh}
\biggl[1-\frac{{\mathcal I}\est(r_w)}{\est^2}
\biggr]
-1\biggr)\,\,-\frac{2\g}{\rn}+C
\,.
\end{equation}
Taking the limit $\rn\rightarrow r_w$, we find

\begin{equation}\label{difference2}
	F_{\QQb}-\sigma\ell=2\g\int_{0}^{r_w} \frac{dr}{r^2}\biggl(\ep^{\s r^2}
\biggl[1-\frac{{\mathcal I}^2_w}{\est^2}\,\biggr]^{\oh}
-1\biggr)\,\,-\frac{2\g}{r_w}+C
\,,
\end{equation}
where ${\mathcal I}_w=\est(r_w)$. Therefore, the asymptotic behavior for large separations takes the form

\begin{equation}\label{FQQb-large}
F_{\QQb}(\ell)=\sigma\ell -2\g\sqrt{\s}\,I+C+o(1)
\,,
\end{equation}
with a dimensionless coefficient
\begin{equation}\label{c}
I=\frac{1}{\sqrt{\s}}\int_{0}^{r_w}\frac{dr}{r^2}
\biggl(1-\ep^{\s r^2}
\biggl[\,1-\frac{{\mathcal I}_w^2}{\est^2}\,\biggr]^{\oh}\biggr)\,\,+\frac{1}{\sqrt{\s}r_w}
\,.
\end{equation}

We conclude our discussion of the connected configuration with some remarks. First, although the world-sheet action includes the two $S_{\text{\tiny A}}$'s associated with the string endpoints, they cancel each other out because the quark and antiquark have opposite $U(1)$ charge. Second, it is worth noting that the constant terms in the expansions of $F_{\QQb}$ are different for small and large $\ell$. Each of those is scheme-dependent, but their difference is not. This makes the model distinct from the simple phenomenological laws like the Cornell model \cite{cornell}. Third, when $T=\mu=0$, $F_{\QQb}$ reduces to the corresponding contribution to the energy of the heavy quark pair.\footnote{It reduces to the energy of the pair only in the absence of dynamical quarks.} It behaves for large interquark separations as 

\begin{equation}\label{EQQb}
E_{\QQb}(\ell)=\sigma_0\ell -2\g\sqrt{\s}\,I_0+C+o(1)
\,,
\end{equation}
where 
\begin{equation}\label{I0}
\sigma_0=\sigma\vert_{T=\mu=0}=\ep\g\s
\,,\qquad
I_0=I\vert_{T=\mu=0}=
\int_0^1\frac{dx}{x^2}\biggl(1+x^2-\ep^{x^2}\Bigl[1-x^4\ep^{2(1-x^2)}\Bigr]^{\frac{1}{2}}\biggr)
\,.
\end{equation}
The above integral is not solvable analytically, whereas a simple numerical calculation gives $I_0\approx 0.751$. 

\subsubsection{Disconnected configurations}
The string configurations $(b)\text{-}(c')$ provide a description of single heavy quarks and antiquarks in the medium. In these cases, the strings are stretched between the horizon and quark sources located on the boundary. Such configurations matter for the computation of the free energies of single particles \cite{a-pol,a-pol2}.\footnote{Although at zero temperature and chemical potential the free energies are infinite, at non-zero temperature and chemical potential they are finite, as it is in QCD with light flavors.} 

First, let us consider configuration $(b)$. It is straightforward to generalize the analysis of \cite{a-pol} to the case of finite chemical potential. The only modifications of the formulas are the obvious $\mu$-dependent terms coming from the boundary action $S_{\text{\tiny A}}\vert_{r=0}$. Indeed, in static gauge the total action is simply

\begin{equation}\label{action-b}
S=S_{\text{\tiny NG}}-\frac{\mu_q}{T}
\,,
\end{equation}
with $\mu_q=\frac{\mu}{3}$, and the analysis then proceeds in the same way as in \cite{a-pol}. As a result, the free energy is given by 

\begin{equation}\label{FQ}
F_{\Q}=\g\sqrt{\s}\,{\cal Q}(h)-\mu_q+\oh C
\,,
\end{equation}
where $h=\s\rh^2$ and the function ${\cal Q}$ is defined by Eq.\eqref{Q}. 

Similarly, for configuration $(b')$ it is 

\begin{equation}\label{FQb}
F_{\Qb}=\g\sqrt{\s}\,{\cal Q}(h)+\mu_q+\oh C
\,.
\end{equation}
This is so because the only modification is due to the sign reversal of $S_A$. It is perhaps noteworthy that for the background geometry \eqref{metric} the free energies can be computed analytically and the results are independent of the form of the blackening factor $f$.

Something similar happens in the two remaining cases. For configuration $(c)$, the total action reduces to 

\begin{equation}\label{action-c}
S=\sum_{i=1}^3 S_{\text{\tiny NG}}\,
+S_{\text{vert}}-\frac{\mu_q}{T}
\,
\end{equation}
so that the rest of the analysis proceeds along the lines of Ref.\cite{a-pol2}. Varying the action with respect to $\rv$ gives

\begin{equation}\label{fb-v}
1+3\k
\Bigl(1+4v-v\frac{\dot f}{f}\,\Bigr)\sqrt{f}\ep^{-3v}=0
\,,
\end{equation}
with $\k=\frac{\tau_v}{3\g}$ and $v=\s\rv^2$. A dot denotes a derivative with respect to $v$. This is a force balance equation at the baryon vertex that determines its position. The expression for the free energy can be written as 

\begin{equation}\label{FQ-V}
F_{\Q'}=\g\sqrt{\s}\Bigl(
2{\cal Q}(h)-{\cal Q}(v)+3\k{\cal V}(f;-2,v)
\Bigr)
-\mu_q+\oh C
\,,
\end{equation}
upon performing the integrals over $r$. The function ${\cal V}$ is defined by \eqref{Vf}.

A similar treatment can be given for configuration $(c')$ which is obtained by replacing $Q$ by $\bar Q$. So, we have

\begin{equation}\label{FQb-V}
F_{\Qb'}=\g\sqrt{\s}\Bigl(
2{\cal Q}(h)-{\cal Q}(v)+3\k{\cal V}(f;-2,v)
\Bigr)
+\mu_q+\oh C
\,.
\end{equation}
We are unable to solve equation \eqref{fb-v} analytically even for the simplest choice of $f$. This significantly complicates the use of the formulas for $F_{\Q'}$ and $F_{\Qb'}$. However, the corresponding analysis can still be done numerically.

%____________________________________________________________________________________
\subsection{Configurations with light quarks}
Here we consider in more detail the basic configurations $(d)\text{-}(e')$. In the context of the gauge/string duality these describe heavy-light mesons and baryons. Unlike \cite{KKW}, we do not introduce probe flavor branes associated with light dynamical quarks, instead we model the disconnected string configurations by assuming the constant tachyon background. Such a background can be interpreted as a source of point-like objects attached to string endpoints in the bulk.

\subsubsection{Configurations without baryon vertices}

First, let us consider configuration $(d)$. Since we are interested in static configurations, we choose the static gauge $\xi^1=t$ and $\xi^2=r$. For the geometry \eqref{metric}, the Nambu-Goto action is then 

\begin{equation}\label{NG1}
S_{\text{\tiny NG}}=\frac{\g}{T}\int dr\,\frac{\ep^{\s r^2}}{r^2}\sqrt{1+f(\partial_r x^i)^2} 
\,.
\end{equation}
From this, it follows that $x^i=const$, which represents a straight string stretched along the $r$-axis, is a solution to the equations of motion. Taking account also of the boundary terms \eqref{Sq} and \eqref{SA}, the total action 

\begin{equation}\label{action-d0}
S=S_{\text{\tiny NG}}+(S_{\text{q}}+S_{\text{\tiny A}})\vert_{r=r_{\bar q}}	+S_{\text{\tiny A}}\vert_{r=0}
\,,
\end{equation}
evaluated on this solution, takes the form

\begin{equation}\label{action-d}
S=\frac{1}{T}\Bigl[
\g\int_0^{\rqb}\frac{dr}{r^2}\ep^{\s r^2}
+
\m\frac{\ep^{\frac{\s}{2}\rqb^2}}{\rqb}\sqrt{f}
+
\frac{1}{3}A_0(\rqb)-\mu_q
\Bigr]
\,.
\end{equation}

Next we extremize the action with respect to the position of the light antiquark. The evaluation of the derivatives just gives

\begin{equation}\label{fb-qb}
\g\ep^{\bar q}+\m\Bigl({\bar q}-1+\bar q\frac{f'}{f}\Bigr)\sqrt{f}\ep^{\oh\bar q}+\frac{2}{3}\sqrt{\frac{{\bar q}^3}{\s}}A'_0=0
\,,
\end{equation}
where $\bar q=\s\rqb^2$. A prime stands for a derivative with respect to $\bar q$. The physical interpretation of \eqref{fb-qb} is clear. This is an equation of force balance at the string endpoint $r=\rqb$. It determines $\bar q$ and can be solved numerically for a particular set of parameters.

From \eqref{action-d}, we can read off the free energy of the configuration 

\begin{equation}\label{FQqb1}
F_{\Qqb}=
\g\int_0^{\rqb}\frac{dr}{r^2}\ep^{\s r^2}
+
\m\frac{\ep^{\frac{\s}{2}\rqb^2}}{\rqb}\sqrt{f}
+
\frac{1}{3}A_0(\rqb)-\mu_q
\,. 
\end{equation}
 The first term is singular, and therefore it requires regularization. As usual, we implement this by imposing a cutoff $\epsilon$ on the lower limit of integration

\begin{equation}\label{regularization}
\int_{\epsilon}^{\rqb}\frac{dr}{r^2}\ep^{\s r^2}=\frac{1}{\epsilon}-\frac{1}{\rqb}+\int_{\epsilon}^{\rqb}\frac{dr}{r^2}\Bigl(\ep^{\s r^2}-1\Bigr)
\,.
\end{equation}
Then subtracting the $\frac{1}{\epsilon}$ term and letting $\epsilon=0$, we get a renormalized free energy 

\begin{equation}\label{FQqb2}
F_{\Qqb}=
\g\int_0^{\rqb}\frac{dr}{r^2}\Bigl(\ep^{\s r^2}-1\Bigr)\,
-\frac{\g}{\rqb}
+\oh C
+
\m\frac{\ep^{\frac{\s}{2}\rqb^2}}{\rqb}\sqrt{f}
+
\frac{1}{3}A_0(\rqb)-\mu_q
\,.
\end{equation}
Here the normalization constant $C$ is the same as in the previous examples. The integral is easily evaluated (see Appendix A), with the result

\begin{equation}\label{FQqb}
F_{\Qqb}=\sqrt{\s}
\Bigl(\g{\cal Q}(\bar q)
+\m{\cal V}(f;\tfrac{1}{2},\bar q)
\Bigr)
+
\frac{1}{3}A_0(\bar q)-\mu_q
+\oh C
\,.
\end{equation}
Thus, the free energy is determined from two equations: \eqref{fb-qb} and \eqref{FQqb}. 

It is straightforward to extend the above analysis to configuration $(d')$. The only modification arises from the sign reversal in the $S_{\text A}$'s. In this case Eq.\eqref{fb-qb} is replaced by 

\begin{equation}\label{fb-q}
\g\ep^{q}+\m\Bigl(q-1+q\frac{f'}{f}\Bigr)\sqrt{f}\ep^{\oh q}-\frac{2}{3}\sqrt{\frac{q^3}{\s}}A'_0=0
\,,
\end{equation}
and Eq.\eqref{FQqb} by 

\begin{equation}\label{FQbq}
F_{\qQb}=\sqrt{\s}\Bigl(
\g{\cal Q}(q)
+\m{\cal V}(f;\tfrac{1}{2},q)
\Bigr)
-
\frac{1}{3}A_0(q)+\mu_q
+\oh C
\,,
\end{equation}
with $q=\s\rq^2$. As a result, the free energy of this configuration is determined by Eqs.\eqref{fb-q} and \eqref{FQbq}. Again, the non-linear equation which determines the position of the string endpoint in the radial direction can be solved only numerically. 

\subsubsection{Configurations with baryon vertices}

The description of configurations $(e)$ and $(e')$ differs only in one way from what we have just described. The new feature is an inclusion of the baryon vertices. 

We begin with configuration $(e)$. It is governed by the following action 

\begin{equation}\label{action-e0}
S=\sum_{i=1}^3 S_{\text{\tiny NG}}+\sum_{i=1}^2(S_{\text{q}}+S_{\text{\tiny A}})\vert_{r=r_{q}}	+S_{\text{\tiny A}}\vert_{r=0}
+S_{\text{vert}}
\,.
\end{equation}
In static gauge, when evaluated on the solutions $x^i=const$, $S$ takes the form 

\begin{equation}\label{action-e}
S=\frac{1}{T}\Bigl[
\g\int_0^{\rv}\frac{dr}{r^2}\ep^{\s r^2}
+
2\g\int_{\rv}^{\rq}\frac{dr}{r^2}\ep^{\s r^2}
+
2\m\sqrt{f}\,\frac{\ep^{\frac{\s}{2}\rq^2}}{\rq}
-
\frac{2}{3}A_0(\rq)-\mu_q
+\tau_v\sqrt{f}\,\frac{\ep^{-2\s\rv^2}}{\rv}
\Bigr]
\,.
\end{equation}
Now it is easy to see that varying $S$ with respect to $\rv$ gives rise to Eq.\eqref{fb-v}, whereas with respect to $\rq$ to Eq.\eqref{fb-q}. Those equations determine $v$ and $q$, respectively. 

From \eqref{action-e}, it follows that the free energy of the configuration is 

\begin{equation}\label{FQqq1}
F_{\Qqq}=
\g\int_0^{\rv}\frac{dr}{r^2}\ep^{\s r^2}
+
2\g\int_{\rv}^{\rq}\frac{dr}{r^2}\ep^{\s r^2}
+
2\m\sqrt{f}\,\frac{\ep^{\frac{\s}{2}\rq^2}}{\rq}
-
\frac{2}{3}A_0(\rq)-\mu_q
+\tau_v\sqrt{f}\,\frac{\ep^{-2\s\rv^2}}{\rv}
\,.
\end{equation}
The first term here is singular. A proper treatment of singularity proceeds in the same way as before and gives 

\begin{equation}\label{FQqq2}
F_{\Qqq}=
\g\int_0^{\rv}\frac{dr}{r^2}\Bigl(\ep^{\s r^2}-1\Bigr)\,
-\frac{\g}{\rv}
+\oh C
+
2\g\int_{\rv}^{\rq}\frac{dr}{r^2}\ep^{\s r^2}
+
2\m\sqrt{f}\,\frac{\ep^{\frac{\s}{2}\rq^2}}{\rq}
-
\frac{2}{3}A_0(\rq)-\mu_q
+\tau_v\sqrt{f}\,\frac{\ep^{-2\s\rv^2}}{\rv}
\,. 
\end{equation}
After evaluating the integrals with the help of the formulas listed in Appendix A, one finds that 

\begin{equation}\label{FQqq}
F_{\Qqq}=\sqrt{\s}
\Bigl(2\g{\cal Q}(q)-\g{\cal Q}(v)
+
2\m{\cal V}(f;\tfrac{1}{2},q)
+
3\g\k{\cal V}(f;-2,v)
\Bigr)
-
\frac{2}{3}A_0(q)-\mu_q
+\oh C
\,, 
\end{equation}
with $\k$ defined by \eqref{fb-v}. Thus, the free energy of configuration $(e)$ is determined by Eqs.\eqref{fb-v}, \eqref{fb-q} and \eqref{FQqq}.

It is clear how to get from the above expression to that for the free energy of configuration $(e')$. All that really matters is the sign reversal in the $S_{\text A}$'s. So, we have 

\begin{equation}\label{FQqqb}
F_{\Qqqb}=
\sqrt{\s}\Bigl(2\g{\cal Q}(\bar q)
-\g{\cal Q}(v)
+
2\m{\cal V}(f;\tfrac{1}{2},\bar q)
+
3\g\k{\cal V}(f;-2,v)
\Bigr)
+
\frac{2}{3}A_0(\bar q)+\mu_q
+\oh C
\,. 
\end{equation}
Here $v$ and $\bar q$ are determined from equations \eqref{fb-v} and \eqref{fb-qb}, respectively.

%__________________________________________________________________
\section{Some details on the model of Sec.III}
%\label{notation}
\renewcommand{\theequation}{C.\arabic{equation}}
\setcounter{equation}{0}

\begin{figure}[H]
\centering
\includegraphics[width=6.5cm]{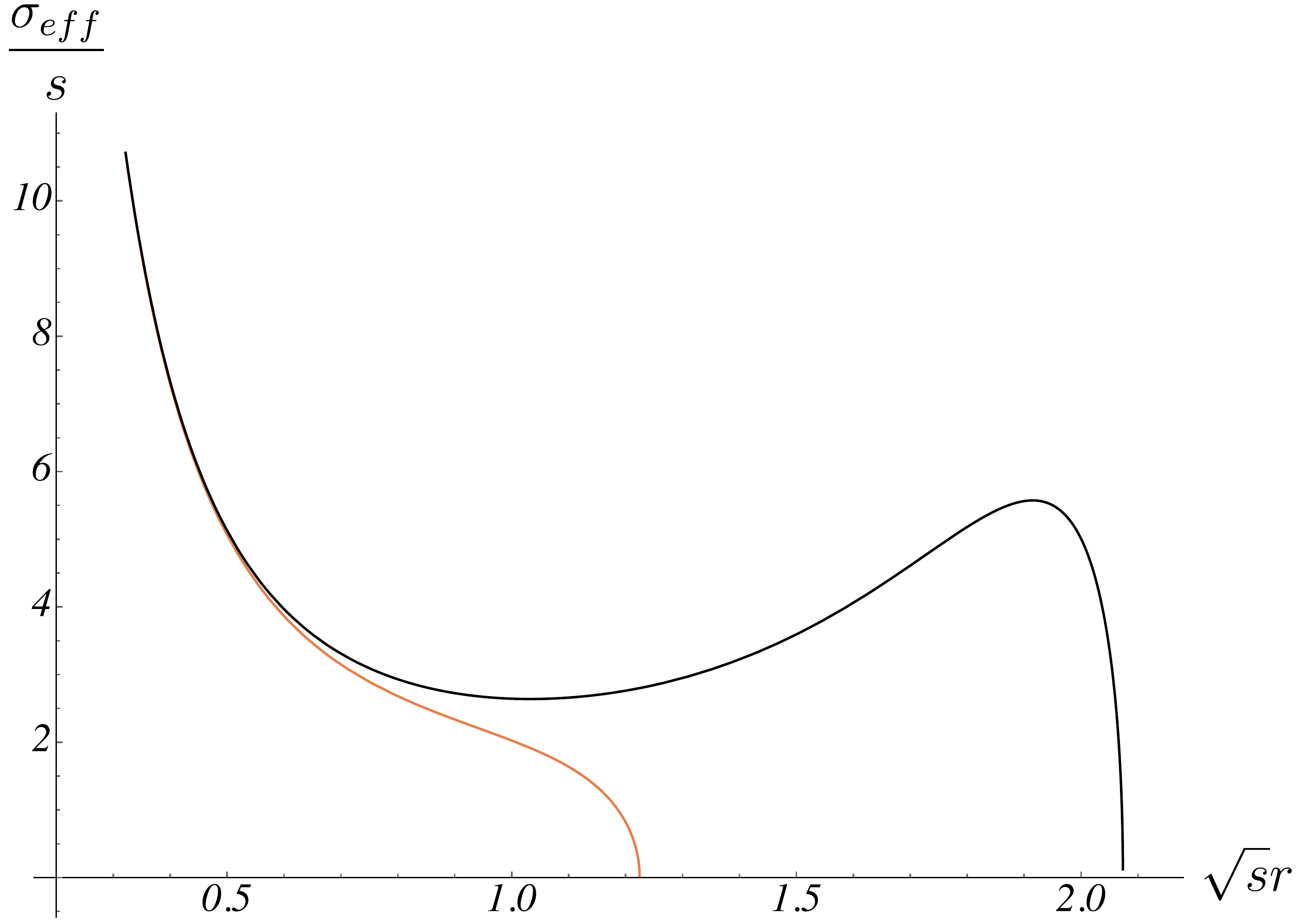}
\caption{{\small Schematic representation of the effective string tension in two different regimes which we call the hadronic phase (upper curve) and the quark-gluon phase (lower curve). Here $r$ runs from $0$ to $\rh$, and $\est(\rh)=0$.}}
\label{Est}
\end{figure}

In the model we are considering the free energy $F_{\QQb}(\ell)$ shows a linear behavior at large quark separations for some temperatures and chemical potentials. Following \cite{az3}, we will explain how to analyze this properly. An important fact is that the effective string tension \eqref{est}, as a function of $r$, could have a local minimum at $r=r_-$ in the interval $(0,\rh)$. If so, then the minimum determines the so-called soft-wall which gives rise to the linear behavior. In the opposite situation the linear behavior does not arise. This is illustrated in Figure \ref{Est}. 

%________________________  f - 8 __________________________________
%______________________________________________________________

We now explore this question for the background \eqref{fnr}. In this case the extrema of $\est(r)$ are determined by solving a quartic equation 

\begin{equation}\label{extrema}
2p^2ht^4+\bigl[p^2-\bigl(1+2p^2\bigr)h\bigr]t^3+ht-1=0
\,,	
\end{equation}
where $t=\frac{r^2}{\rh^2}$. For given $p$, this equation has two roots in the interval $(0,1)$, if $h>h_p$,  where $h_p$ is a solution to an algebraic equation in the variable $h$

\begin{equation}\label{critline}
12p^4f_1^2\biggl(\frac{f_2}{f_3}+\frac{f_3}{12p^4h^2}\biggr)^2
+
8p^4\biggl(\frac{f_2}{f_3}+\frac{f_3}{12p^4h^2}\biggr)^3
-4p^2f_1^3+1=0
\,.
\end{equation}
Here
\begin{equation}\label{f13}
f_1=\frac{1}{4p^2h}\Bigl(\bigl(1+2p^2\bigr)h-p^2\Bigr)
\,,\quad
f_2=\bigl(1+2p^2\bigr)h^2-9p^2h
\,,\quad
f_3=6\sqrt[3]{p^8h^5}
\biggl[h-8p^2f_1^2+\biggl(\bigl(h-8p^2f_1^2\bigr)^2-\frac{f_2^3}{27p^4h^4}\biggr)^{\frac{1}{2}}\biggr]^{\frac{1}{3}}
\,.
\end{equation}
In this case the smaller and larger roots, which respectively corresponds to the minimum and maximum of $\est$, are given by\footnote{At $h=h_p$, the roots are equal to each other.} 

\begin{equation}\label{rmin}
t_{\mp}=\oh\,
\Biggl[f_1
\mp
\biggl(f_1^2+\frac{f_2}{f_3}+\frac{f_3}{12p^4h^2}\biggr)^{\frac{1}{2}}
\pm
\Biggl(2f_1^2-\frac{f_2}{f_3}-\frac{f_3}{12p^4h^2}
\pm
\frac{1-2p^2f_1^3}{p^2\Bigl(f_1^2+\frac{f_2}{f_3}+\frac{f_3}{12p^4h^2}\Bigr)^{\frac{1}{2}}}\Biggr)^{\frac{1}{2}}\,
\Biggr]
\,.\qquad
\end{equation}
Note that in the limit $h\rightarrow\infty$ the smaller root $r_-$ goes to $1/\sqrt{\s}$, as expected \cite{az1}.

There are two special cases: $p=0$ and $p=1$, which represent the cases of zero chemical potential and temperature, respectively. First, let us specialize to the case $p=0$. It is easy to see that in this case Eq.\eqref{extrema} reduces to a cubic equation 

\begin{equation}\label{extrema0}
ht^3-ht+1=0
\,. 
\end{equation}
In the interval $(0,1)$ this equation has two roots \cite{az3}

\begin{equation}\label{rminmu=0}
t_-=\frac{2}{\sqrt{3}}\sin\Bigl(\frac{1}{3}\arcsin\frac{h_0}{h}\Bigr)
\\,\qquad
t_+=\frac{2}{\sqrt{3}}\sin\Bigl(\frac{\pi}{3}-
\frac{1}{3}\arcsin\frac{h_0}{h}\Bigr)
\,,
\end{equation}
if $h>h_0$, where 
\begin{equation}\label{hc0}
h_0=\frac{3\sqrt{3}}{2}\approx 2.60
\,.	
\end{equation}

Now let us move on to the second case. At $p=1$, the left hand side of Eq.\eqref{extrema} is factorized

\begin{equation}\label{extrema1}
\bigl(t-1\bigr)\bigl(2ht^3+(1-h)t(t+1)+1\bigr)=0
\,. 
\end{equation}
We may omit the first factor, since we are only interested in the interval $(0,1)$. Thus, the original equation simplifies and becomes a cubic one 

\begin{equation}\label{extrema11}
2ht^3+(1-h)t^2+(1-h)t+1=0
\,. 
\end{equation}
In the interval $(0,1)$ it has two roots 

\begin{equation}\label{rminT=0}
\begin{split}
t_-=&\frac{h-1}{6h}+\frac{1}{3h}\sqrt{7h^2-8h+1}\sin\Bigl(\frac{1}{3}
\arcsin\frac{10h^3-75h^2+12h-1}{(7h^2-8h+1)^{\frac{3}{2}}}\Bigr)
\,,\\
t_+=&\frac{h-1}{6h}+\frac{1}{3h}\sqrt{7h^2-8h+1}\cos\Bigl(\frac{1}{3}
\arccos\frac{10h^3-75h^2+12h-1}{(7h^2-8h+1)^{\frac{3}{2}}}\Bigr)
\,,
\end{split}
\end{equation}
if $h>h_1$, where 

\begin{equation}\label{hc1}
h_1=\frac{1}{3}\Bigl(\sqrt{84-\Delta+110\Delta^{-\oh}}-1-\Delta^{\oh}\Bigr)
\,,\qquad
	\Delta=28-57\Bigl(\frac{169-15\sqrt{5}}{2}\Bigr)^{-\frac{1}{3}}-3\Bigl(\frac{169-15\sqrt{5}}{2}\Bigr)^{\frac{1}{3}}
	\,.
\end{equation}
The value of $h_1$ is approximately $3.50$. 

%__________________       R E F s     ______________________

%____________________________________________________________________
\end{document}